\documentclass{article}
\usepackage[utf8]{inputenc} 
\usepackage[T1]{fontenc}    
\usepackage{hyperref}       
\usepackage{url}            
\usepackage{booktabs}       
\usepackage{nicefrac}       
\usepackage{microtype}      
\usepackage{amsfonts}       
\usepackage{amsmath}
\usepackage{amssymb}
\usepackage{paralist}
\usepackage{times}
\usepackage{graphicx,subfig}
\usepackage{graphics}
\usepackage{tabularx}
\usepackage{amsfonts,amsmath,amssymb,amsthm,url,xspace}
\usepackage{geometry}
\usepackage[numbers]{natbib}
\usepackage{algorithm}
\usepackage{algorithmic}
\usepackage{lipsum}
\usepackage{setspace}
\usepackage{verbatim}
\usepackage{pifont}
\usepackage{authblk}
\usepackage{mathrsfs}

\usepackage[colorinlistoftodos]{todonotes}
\newtheorem{lem}{Lemma}
\newtheorem{thm}{Theorem}

\def\b0{{\boldsymbol{0}}}
\def\bd{{\boldsymbol{d}}}

\def\bx{{\boldsymbol{x}}}
\def\blambda{{\boldsymbol{\lambda}}}

\def\br{{\boldsymbol{r}}}

\def\bv{{\boldsymbol{v}}}
\def\bu{{\boldsymbol{u}}}
\def\ba{{\boldsymbol{a}}}

\def\bb{{\boldsymbol{b}}}

\def\by{{\boldsymbol{y}}}

\def\CDdual-ls{{\sf CDdual-ls}\xspace}

\def\CDdual{{\sf CDdual}\xspace}

\def\R{\mathfrak{R}}
\def\X{\mathcal{X}}
\def\Y{\mathcal{Y}}

\def\CCDR1{{\sf CCD++}\xspace}

\def\e2006{{\sf TFIDF-2006}\xspace}

\def\rcv1{{\sf rcv1}\xspace}

\def\a9a{{\sf a9a}\xspace}

\def\ml1m{{\sf movielens1m}\xspace}

\def\movielens10m{{\sf movielens10m}\xspace}

\def\x{x}

\newcommand{\bw}{{\boldsymbol{w}}}

\newcommand{\bepsilon}{{\boldsymbol{\epsilon}}}

\title{Sparse Learning with Semi-Proximal-Based Strictly Contractive Peaceman-Rachford Splitting Method}
\author[1]{Sen Na\thanks{senna@uchicago.edu}}
\author[2]{Cho-Jui Hsieh\thanks{chohsieh@ucdavis.edu}}
\affil[1]{Department of Statistics, The University of Chicago}
\affil[2]{Department of Statistics, University of California, Davis}

\date{\vspace{-5ex}}

\begin{document}	
	\maketitle

\begin{abstract}
In this paper, we will focus on solving the splitting problem which is minimizing the sum of two convex functions subject to a linear constraint. This problem has attracted tremendous attention because of its wide applications to machine learning problems, such as Lasso, group Lasso, Logistic regression and image processing. A recent paper by Gu et al (2015) developed a Semi-Proximal-Based Strictly Contractive Peaceman-Rachford Splitting Method (SPB-SCPRSM), which is an extension of Strictly Contractive Peaceman-Rachford Splitting Method (SCPRSM) proposed by He et al (2014). By introducing semi-proximal terms and using two different relaxation factors, SPB-SCPRSM showed a more flexiable applicability compared with its origin SCPRSM and widely-used Alternating Direction Method of Multipliers (ADMM) algorithm, though all of them have $O(1/t)$ convergence rate. 
In this paper, we develop a stochastic version of SPB-SCPRSM, where only a subset of samples (or even only one sample) are used at each iteration.
The resulting algorithm (Stochastic SPB-SCPRSM) can not only scale to problems with huge number of samples and also be more flexiable than Stochastic ADMM on the numerical experiment by setting semi-proximal terms and relaxation factors. Moreover, we show that our proposed method has $O(1/\sqrt{t})$ convergence rate in ergodic sense in general and $O(\log(t)/t)$ in strong convexity case.
%
\end{abstract}

\section{Introduction}

In this paper, we mainly consider the convex minimization problem with separable objective functions and a linear constraint. The problem can be formulated as:
\begin{equation}
  \min_{\bx \in \X, \by \in \Y} \theta_1(\bx) + \theta_2(\by) \ \text{ s.t. } A\bx+B\by = \bb,
  \label{eq:batch}
\end{equation}
where $\theta_1: \R^{d_1}\rightarrow \R$ and $\theta_2: \R^{d_2}\rightarrow \R$ are convex functions and $\X, \Y$  are nonempty convex sets. Typically, $\theta_1(\bx)={1 \over n} \sum^n_{i=1}\theta_{1i}(\bx)$, where
 $n$ is the number of observations and $\theta_{1i}(\bx)$ is the convex loss incurred on observation $i$, and $\theta_2(\by)$ is the structural regularization term.

%

This minimization problem can be solved by a group of splitting algorithms. The Alternating Direction Method of Multipliers (ADMM) \cite{GM75, GM76, SB11a} proposed in 1970s is one of the simplest methods. It provides a flexible framework to handle each component individually. As analyzed in the previous paper \cite{G83}, ADMM is equivalent to applying Douglas-Rachford Splitting Method (DRSM)~\cite{DR56, LM79} to the dual problem of (1). Considerable researches have been conducted in the recent 30 years to analyze the convergence properties of ADMM~\cite{G83,GT89,EB92}.

At the same time, the Peaceman-Rachford Splitting Method (PRSM)
~\cite{PR55, LM79} has also been proposed to solve the batch splitting problem~\eqref{eq:batch}. It often converges faster than ADMM, but requires more restrictive assumptions to ensure the convergence. Recently, a modified version called Strictly Contractive PRSM (SCPRSM) was developed by He et al. (2014). They relax the requirements of PRSM for convergence by employing a suitable relaxation factor (usually between (0, 1)), that makes PRSM converge under the same assumption with ADMM. 
Additionally, Semi-Proximal-Based SCPRSM, as an extension of SCPRSM, was also proposed by Gu et al. (2015) ~\cite{YBD15}. They introduced two semi-proximal terms in the iteration scheme of SCPRSM with two different relaxation factors to make it more flexible. To be clarified, all the above algorithms have $O(1/t)$ convergence rate in general but it was reported in He et al. (2014) that PRSM based algorithms are usually faster than ADMM on many synthetic and real datasets.

However, when the dataset has a large sample size (which may not fit in a single machine), all the above methods cannot scale well because they adopt the "batch" setting, which means they need to \textit{visit all the samples at each iteration}. As a consequence, regradless of ADMM, PRSM, SCPRSM, or SPB-SCPRSM, they are all not suitable for big data applications. To alleviate this problem, on the other hand, a family of stochastic ADMM algorithms have been proposed~\cite{OH13,WB12,TZ13}, where only one or a mini-batch of samples are used at each iteration. Due to the scalability to large datasets, stochastic ADMM algorithms have become a popular research topic~\cite{OH13,WB12,TZ13, LWZ14a,SA14a}. Comparing to batch algorithm, consuming time for each iteration is much less in the stochastic algorithm though it also loses some information. So, convergence rate for stochasic algorithm is usually under ergodic sense (only hope the mean is as close to true value as possible). 

In this paper, we propose a stochastic version of the SPB-SCPRSM algorithm. The resulting algorithm, Stochastic SPB-SPRSM, only requires one or a small subset of training samples at each time. As a consequence, our algorithm can be easily scaled to problems with a large number of training data and the convergence performance is also better than previous Stochastic ADMM and other related algorithms.
Our contribution can be summarized as follows:
\begin{enumerate}
  \item We extend the batch SPB-SCPRSM to the stochastic setting, where we use the first order approximated Lagrangian to get the "approximated" dual problem and further derive the final algorithm. Our algorithm, Stochastic SPB-SCPRSM, is useful in the following two cases: (1) problem with large sample size as well as high dimensional model parameters, (2) the proximal mapping of loss function (smooth part) cannot be solved efficiently (i.e. subproblem is hard to be solved).
  \item As analyzed in the previous paper ~\cite{YBD15}, we also add two proximal terms onto subproblems for updating the $\bx$ and $\by$ and use two different relaxation factors to make our algorithm more flexible. We prove the bound of them for make the iteration sequence generated by Stochastic SPB-SPRSM be strictly contractive is the same as SPB-SPRSM:

      $$\alpha \in \left[0,1\right),\gamma \in \left(0,\frac{1-\alpha+\sqrt{(1+\alpha)^2+4(1-\alpha^2)}}{2}\right)$$,

      where $\alpha$ and $\gamma$ are two relaxation factors.\\
      \textit{Note}: we can get several stochastic algorithms like Stochastic ADMM and Stochastic SCPRSM by setting different semi-proximal terms and relaxation factors.
  \item We show that the convergence rate of the proposed algorithm is $O(1/\sqrt{t})$ in the ergodic sense under the same assumptions with Stochastic ADMM and $O(\log(t)/t)$ in the strong convexity case. So, we also unify how to analyze the convergence rate of Stochastic SPRSM ($\alpha=\gamma$) and Stochastic ADMM ($\alpha=\gamma$ goes to 1).
  \item We conduct experimental comparisons with Stochastic ADMM algorithms and other related algorithms on several synthetic and real datasets.
  \end{enumerate}

  We begin by presenting the background of splitting problem in Section~\ref{sec:background}. In Section~\ref{sec:algorithm},  we propose the iteration of our main algorithm (Stochastic SPB-SPRSM). The main results of convergence rate is provided in Section~\ref{sec:conv}. The experimental results on simulated and real datasets are presented in Section~\ref{sec:4}. Finally, we give conclusions as well as future possible work in Section~\ref{sec:summary}. Our proofs are included in the appendix.

\section{Background}
\label{sec:background}

In this section, we first present the stochastic optimization
problem we want to solve, and then discuss
several related algorithms for both batch and stochastic settings.

\subsection{Stochastic Setting}
We mainly study the convex stochastic optimization problem of the form:
\begin{equation}
\label{eq:1-1}
\min_{\bx\in\X, \by\in\Y} E_{\xi} \theta_1(\bx, \xi) + \theta_2(\by)
  \text{ s.t. }  A\bx+B\by = \bb,
\end{equation}
where $\bx\in\mathfrak{R}^{d_1}, \by\in\mathfrak{R}^{d_2},A\in\mathfrak{R}^{m\times d_1},B\in\mathfrak{R}^{m\times d_2},\bb\in\mathfrak{R}^{m}$; $\theta_1(\bx,\xi )$ is the instance function value while $\theta_1(\bx)=E_\xi\theta_1(\bx,\xi )$ is its expectation; $\theta_2: \mathfrak{R}^{d_2}\rightarrow\mathfrak{R}$ is a composite function. Both $\theta_1$ and $\theta_2$ are convex functions but can be nonsmooth,  and $\X$ and $\Y$ are closed convex set.
The random variable $\xi$ follows some fixed but unknown distribution $P$, and we are able to draw a sequence of i.i.d samples from this distribution.
When $\xi$ is deterministic, we can recover the original problem formulation~\eqref{eq:batch}, and the algorithm
also becomes deterministic (batch) algorithm.

Problem~\eqref{eq:1-1} covers the following structural risk minimization problems in machine learning:
\begin{equation*}
\min_{\bx} E_{\xi} L(\bx, \xi) + R(\bx),
\end{equation*}
where $\bx$ is the model parameter, $L(\cdot, \xi)$ is the loss function on a sample $\xi$, and $R(\cdot)$
is the regularizer that imposes some structural constraints. If the training dataset with $n$ samples is given,
the empirical risk minimization problem $\frac{1}{n}\sum_{i=1}^nL(\bx, \xi_i)+R(\bx)$ can also be modeled
this way where $\xi$ is uniformly sampled from training samples.
Note that the structural regularizer can often be non-smooth, throughout this paper we assume $\theta_1, \theta_2$
are convex but may be non-smooth (all assumptions are same with Stochastic ADMM).

The stochastic optimization algorithms for solving~\eqref{eq:1-1} are more suitable for large-scale machine learning problems compared with batch algorithms.
In the stochastic setting, the algorithm only observes one sample or a subset of samples at each time, which is useful when (i) the dataset is larger
than the memory or (ii) the data points come from a streaming model. Therefore, many recent papers~\cite{OH13,WB12,TZ13, LWZ14a,SA14a} discuss Stochastic ADMM algorithm or incremental algorithms for solving problem~\eqref{eq:1-1}.

\subsection{Related Batch Algorithms}
We first discuss the ADMM for solving~\eqref{eq:batch}, which aims to minimize the following augmented Lagrangian:
\begin{align}
  L_{\beta}(\bx, \by, \blambda) &= \theta_1(\bx) + \theta_2(\by) - \langle \blambda, \ A\bx+B\by - \bb\rangle \nonumber
  + \frac{\beta}{2} \|A\bx + B\by - \bb\|^2,
  \label{eq:lagrangian}
\end{align}
where $\blambda\in\R^{m}$ is the Lagrangian multiplier and $\beta>0$ is a penalty parameter.
The ADMM algorithm minimizes $L_\beta$ in a Gauss-Seidel (or block coordinate descent) manner---sequentially updating $\bx, \by, \blambda$
at each iteration, as shown in Algorithm~\ref{alg:1}.
  \begin{algorithm}[tb]
  	\caption{Deterministic ADMM}
  	\label{alg:1}
  	\begin{algorithmic}
  		\STATE Initialize $y_0$ and $\lambda _0=0$.
  		\FOR{$k=0, 1, 2,\ldots$}
  		\STATE $\bx_{k+1}\leftarrow\arg\min _{\bx\in\X}L_\beta (\bx, \by_k; \blambda _k)$.
  		\STATE $\by_{k+1}\leftarrow\arg\min _{\by\in\mathcal{Y}}L_\beta (\bx_{k+1}, \by; \blambda _k)$.
  		\STATE $\blambda _{k+1}\leftarrow\blambda _k-\beta (A\bx_{k+1}+B\by_{k+1}-\bb)$.
  		\ENDFOR
  	\end{algorithmic}
  \end{algorithm}

  Peaceman-Rachford Splitting Method (PRSM) is another way to minimize the Lagrangian~\cite{PR55, LM79}. By adding a step of updating $\blambda_{k+1/2}$ immediately after updating $\bx$, PRSM performs a faster contraction speed in experiment whenever it is convergent, though it does not improve the convergence rate or even becomes less robust (it converges only for some specific problems). 
  To make PRSM converge for general cases, Strictly Contractive PRSM (SCPRSM) was recently proposed in~\cite{HL14}, and applied to many statistical learning models such as Lasso, group Lasso, sparse logistic regression and image processing. The key idea is to employ an underdetermined relaxation factor $\alpha \in \left(0,1\right)$ when updating the Lagrange multipliers $\blambda$ (see Algorithm~\ref{alg:2}).
  \begin{algorithm}[tb]
  	\caption{Strictly Contractive PRSM ($\alpha=1$ for PRSM)}
  	\label{alg:2}
  	\begin{algorithmic}
  		\STATE Initialize $y_0$ and $\lambda _0=0$.
  		\FOR{$k=0, 1, 2,\ldots$}
  		\STATE $\bx_{k+1}\leftarrow\arg\min _{\bx\in\X}L_\beta (\bx, \by_k; \blambda _k)$.
        \STATE $\blambda _{k+1/2}\leftarrow\blambda _k-\alpha\beta (A\bx_{k+1}+B\by_{k}-\bb)$.
  		\STATE $\by_{k+1}\leftarrow\arg\min _{\by\in\mathcal{Y}}L_\beta (\bx_{k+1}, \by; \blambda _{k+1/2})$.
  		\STATE $\blambda _{k+1}\leftarrow\blambda _{k+1/2}-\alpha\beta (A\bx_{k+1}+B\by_{k+1}-\bb)$.
  		\ENDFOR
  	\end{algorithmic}
  \end{algorithm}

  Recently, a new variant of SCPRSM called Semi-Proximal-Based SCPRSM (SPB-SCPRSM) was published ~\cite{YBD15} and it could cover the SCPRSM case. Because the relaxation factor in two iteration steps plays different roles (one is for updating $\by$ and one is for updating $\blambda$), they naturally used two different relaxation factors and also introduced two positive semi-definite matrices in subproblems. All work can make SCPRSM easy to apply (see Algorithm~\ref{alg:3}).
  \begin{algorithm}[tb]
  	\caption{Semi-Proximal-Based SPRSM}
  	\label{alg:3}
  	\begin{algorithmic}
  		\STATE Initialize $y_0$ and $\lambda _0=0$; Suppose $S,T\succcurlyeq 0$.
  		\FOR{$k=0, 1, 2,\ldots$}
  		\STATE $\bx_{k+1}\leftarrow\arg\min _{\bx\in\X}L_\beta (\bx, \by_k; \blambda _k)+{1\over 2} \|\bx-\bx_k\|_S^2$.
        \STATE $\blambda _{k+1/2}\leftarrow\blambda _k-\alpha\beta (A\bx_{k+1}+B\by_{k}-\bb)$.
  		\STATE $\by_{k+1}\leftarrow\arg\min _{\by\in\mathcal{Y}}L_\beta (\bx_{k+1}, \by; \blambda _{k+1/2})+{1\over 2} \|\by-\by_k\|_T^2$.
  		\STATE $\blambda _{k+1}\leftarrow\blambda _{k+1/2}-\gamma\beta (A\bx_{k+1}+B\by_{k+1}-\bb)$.
  		\ENDFOR
  	\end{algorithmic}
  \end{algorithm}

\subsection{Related Stochastic Algorithms}

For making algorithm possible to apply to a large scale dataset, a family of Stochastic ADMM algorithms have been proposed in past 4 years. The pioneer paper~\cite{WB12} presents a splitting algorithm in the online setting, but they only focused on minimizing the online regret. The Stochastic ADMM algorithm~\cite{OH13} first explicitly focused on solving the stochastic version of~\eqref{eq:1-1}.
At each iteration, the algorithm samples a $\xi_{k+1}$ (corresponding to one or a subset of samples), and use this information to update the parameters from $(\bx^{k}, \by^k, \blambda^k)$ to $(\bx^{k+1}, \by^{k+1}, \blambda^{k+1})$.
We will describe the detailed settings in the next section. It was shown
in Ouyang et al. (2013) that the averaged iterates $\bar{\bx}_{t}:=\frac{1}{t}\sum_{i=1}^t \bx_i$, $\bar{\by}_{t}:=\frac{1}{t}\sum_{i=1}^t \by_{i+1}$ have the following
bound:
\begin{align*}
  E\bigg[&\theta_1(\bar{\bx}_t)  + \theta_2(\bar{\by}_t) - \theta_1(\bx^*)-\theta_2(\by^*) +
  \rho\|A\bar{\bx}_t + B\bar{\by}_t - \bb\|_2 \bigg] = O(\frac{1}{\sqrt{t}}),
\end{align*}
where $\bx^*$ and $\by^*$ are optimal solutions.

Due to the application in large-scale problems, many papers~\cite{TZ13, LWZ14a,SA14a} improve the algorithm and/or analysis of Stochastic ADMM within the past two years.

\subsection{Notations}

All notations are consistent throughout the paper and supplementary materials.
We denote the objective function as $\theta(\bu) = \theta_1(\bx) + \theta_2(\by)$; the constraint function as $\br(\bw)=A\bx+B\by-\bb$ and its \emph{k}th iteration as $\br_k=A\bx_k+B\by_k-\bb$; the residual term as $\delta _k=\theta _1'(\bx_{k-1},\xi _k)-\theta _1'(\bx_{k-1})$.
For simplicity, we define the following vectors that will be used in this paper:
\[ \bu=\left( \begin{array}{ccc}
\bx \\
\by  \end{array} \right) ,
 \bv=\left( \begin{array}{ccc}
\by \\
\blambda  \end{array} \right) ,
 \bw=\left( \begin{array}{ccc}
\bx \\
\by \\
\blambda \end{array} \right),
 \bar{\bu}_t=\left( \begin{array}{ccc}
\frac{1}{t}\sum_{k=1}^{t}\bx_k \\
\frac{1}{t}\sum_{k=1}^{t}\by_{k+1}  \end{array} \right),
\bar{\bw}_t=\left( \begin{array}{ccc}
\frac{1}{t}\sum_{k=1}^{t}\bx_{k+1} \\
\frac{1}{t}\sum_{k=1}^{t}\by_{k+1}  \\
\frac{1}{t}\sum_{k=1}^{t}\blambda_{k+1}  \end{array} \right),\]
\[ F(\bw)=\left( \begin{array}{ccc}
-A^T\blambda \\
-B^T\blambda \\
A\bx+B\by-\bb \end{array} \right),
 P=\left( \begin{array}{ccc}
  S & 0 \\
  0 & T \end{array} \right),
   M=\left( \begin{array}{ccc}
  I_{n_2} & 0 \\
  \alpha\beta B & (\alpha+\gamma)\beta I_m \end{array} \right),\]
\[ K=\left( \begin{array}{ccc}
(1-\alpha)\beta B^T B & (1-\alpha)\beta B^T \\
(1-\alpha)\beta B & (2-\alpha-\gamma)\beta I_m \end{array} \right),
H=\frac{1}{\alpha+\gamma}\left( \begin{array}{ccc}
(\alpha+\gamma-\alpha\gamma)\beta B^{T}B & -\alpha B^T \\
-\alpha B & \frac{1}{\beta}I_m \end{array} \right),\]
\[ G=\left( \begin{array}{ccc}
  P & 0 \\
  0 & 0 \end{array} \right)+
  \left( \begin{array}{ccc}
  0 & 0 \\
  0 & H \end{array} \right)
 =\left( \begin{array}{ccc}
  S & 0 &0 \\
  0 & T+\frac{\alpha+\gamma-\alpha\gamma}{\alpha+\gamma}\beta B^{T}B & -\frac{\alpha}{\alpha+\gamma}B^T \\
  0 & -\frac{\alpha}{\alpha+\gamma}B & \frac{1}{(\alpha+\gamma)\beta}I_m \end{array} \right).\]
\\For positive semidefinite matrix $G$, we also define $G$-norm of $\bx$ as $||\bx||_G=\sqrt{\bx^T G\bx}$.
For a function $f$, we use $f'$ to denote the subgradient of $f$  and if $f$ is differentiable, we denote its derivative as $\nabla f$. We also assume the optimal solution exists and define the supremum distance: $D_\X=\sup\{||\bx_a-\bx_b|| | \bx_a,\bx_b \in \X\}$.

\section{Stochastic Semi-Proximal-Based SPRSM}
\label{sec:algorithm}
In the stochastic setting, given $\xi_{k+1}$ sampled from some fixed but unknown distribution P, the augmented Lagrangian can be approximated by:
\begin{align*}
  \hat{L}_{\beta, k}(\bx, \by, \blambda) &:= \theta_1(\bx_k) + \langle \theta_1'(\bx_k, \xi_{k+1}), \bx-\bx_k\rangle +\theta_2(\by)- \langle \blambda , A\bx+B\by-\bb\rangle \\
  &+ \frac{\beta}{2}\|A\bx+B\by-\bb\|^2_2+\frac{1}{2\eta_{k+1}}\|\bx-\bx_k\|^2,
\end{align*}
where $\beta\geq0$ is still the Lagrangian multiplier. We use $\theta_1'$ to denote a subgradient of $\theta _1$ and $\eta _{k+1}$ is the time-varying step size and $\xi_{k+1}$ is sampled from some fixed but unknown distribution $P$.
As we will show in Section \ref{sec:conv}, different choices of $\eta _k$ will lead to different convergence rates.

Comparing to $L_{\beta}$, we approximate $\theta_1(\cdot)$ by its first order approximation
$\theta_1(\bx_k+\Delta)\approx\theta_1(\bx_k)+\langle \theta_1'(\bx_k, \xi_{k+1}),\Delta\rangle + \frac{1}{2\eta_{k+1}}\|\Delta\|^2$,
where the subgradient is evaluated using the current sample $\xi_{k+1}$. This leads to the following two
computational benefits: (i) Only a subset of samples is used at each time, and (ii) the update for $\bx$ can often be computed in closed form even when $\theta_1$ is a complicated loss function (such as the logistic loss).
Based on the definition of saddle point $(\bx^*,\by^*,\blambda^*)$, we have:
\begin{align*}
	\hat{L}_{\beta ,k}(\bx^*,\by^*,\blambda )\leq&\hat{L}_{\beta ,k}(\bx^*,\by^*,\blambda^*)
	\leq\hat{L}_{\beta ,k}(\bx,\by,\blambda^*)\ \    \forall \bx\in \X,\forall \by\in \Y,\forall \blambda\in\Lambda,
\end{align*}
where $\bx^*, \by^*, \blambda^*$ are optimal solutions.

Based on this approximated Lagrangian, we propose a Stochastic SPB-SPRSM algorithm
in Algorithm~\ref{alg:5}. The key idea is to replace step 1 and 3 (in Algorithm~\ref{alg:3}) by minimizing the approximated Lagrangian, which can let us draw i.i.d sample points from observed data and apply our stochastic setting.  
The reason we use the SPB-SCPRSM approach to update the dual variables $\blambda$ is that adding semi-proximal terms in iterations of $\bx$ and $\by$ and using different relaxation factors can make our algorithm more flexible. We will show that our update rule for $\bx$, $\by$, and $\blambda$ leads to a faster convergence speed on both synthetic and real experiments.

\begin{algorithm}[tb]
	\caption{Stochastic ADMM}
	\label{alg:4}
	\begin{algorithmic}
		\STATE Initialize $x_0$, $y_0$ and set $\lambda _0=0$.
		\FOR{$k=0, 1, 2,\ldots$}
		\STATE $\bx_{k+1}\leftarrow\arg\min _{\bx\in\X}\hat{L}_{\beta ,k}(\bx, \by_k, \blambda _k)$.
		\STATE $\by_{k+1}\leftarrow\arg\min _{\by\in\mathcal{Y}}\hat{L}_{\beta ,k}(\bx_{k+1}, \by, \blambda _k)$.
		\STATE $\blambda _{k+1}\leftarrow\blambda _k-\beta (A\bx_{k+1}+B\by_{k+1}-\bb)$.
		\ENDFOR
	\end{algorithmic}
\end{algorithm}

\begin{algorithm}[tb]
	\caption{Stochastic SPB-SCPRSM}
	\label{alg:5}
	\begin{algorithmic}
		\STATE Initialize $x_0$, $y_0$ and set $\lambda _0=0$; Suppose $S,T\succcurlyeq0$.
		\FOR{$k=0, 1, 2,\ldots$}
		\STATE $\bx_{k+1}\leftarrow\arg\min _{\bx\in\X}\hat{L}_{\beta ,k}(\bx, \by_k, \blambda _k)+{1\over 2} \|\bx-\bx_k\|_S^2$.
        \STATE $\blambda _{k+1/2}\leftarrow\blambda _k-\alpha\beta (A\bx_{k+1}+B\by_{k}-\bb)$.
		\STATE $\by_{k+1}\leftarrow\arg\min _{\by\in\mathcal{Y}}\hat{L}_{\beta ,k}(\bx_{k+1}, \by, \blambda _{k+1/2})+{1\over 2} \|\by-\by_k\|_T^2$.
		\STATE $\blambda _{k+1}\leftarrow\blambda _{k+1/2}-\gamma\beta (A\bx_{k+1}+B\by_{k+1}-\bb)$.
		\ENDFOR
	\end{algorithmic}
\end{algorithm}

\section{Convergence Analysis}
\label{sec:conv}

In this section, we will show that our Stochastic SPB-SCPRSM has a rate $O(1 / \sqrt{t})$ of convergence under the same assumptions of Stochastic ADMM. That is $\forall \rho>0$ we have:
\begin{displaymath}
E\bigg[ \theta(\bar{\bu}_t) - \theta(\bu^*) + \rho\|A\bar{\bx}_t+B\bar{\by}_t-\bb\|_2 \bigg] = O(\frac{1}{\sqrt{t}}),
\end{displaymath}
where $\bu^*$ is the optimal solution.
Note that in our algorithm if we set $S=T=0$ and $\alpha=0$, $\gamma=1$, we will get Stochastic ADMM and our convergence conclusions coincide with Stochastic ADMM. If we set $S=T=0$ and $\alpha=\gamma \in (0,1)$, we will get Stochastic SCPRSM. So, we know the convergence rate of Stochastic SCPRSM is also $O(1 / \sqrt{t})$.
From this point, we unify the analysis of convergence rate of several stochastic algorithms. Moreover, if $\theta_1$ is a strongly convex function, we can strength the rate to $O(\log(t)/t)$.
All proofs in this section are provided in the appendix, and we just list the important lemmas and a sketch of the proof here.

\noindent{\bf Assumptions. } To prove the convergence results, we need the following assumptions:
\begin{enumerate}
	\item $\theta_1$ and $\theta_2$ are convex functions but not necessary smooth.
	\item For $\forall \bx \in \X$, we have $E [|| \theta_{1}'(\x_k, \xi_{k+1})||^2 ] \leq N^2$, where $N$ is a constant.
\end{enumerate}

\noindent{\bf Proof Sketch. }
In the following, we list four important lemmas for getting the final theorem.
\begin{lem}
	\label{lem:1}
	Let $\bw_{k}=(\bx_{k}, \by_{k}, \blambda_{k})$ be the sequence generated by the iteration scheme of the Stochastic SPB-SCPRSM (Algorithm~\ref{alg:5}).
	If we define $d_k=\|\bx-\bx_k\|^2-\|\bx-\bx_{k+1}\|^2$ and \begin{align*}
	P_1&=\theta(\bu)-(\theta_1(\bx_k)+\theta_2(\by_{k+1}))+\langle \bx-\bx_k,\delta_{k+1}\rangle+\frac{\eta_{k+1}}{2} \|\theta_1'(\bx_k,\xi_{k+1})\|^2
	+\frac{d_k}{2\eta_{k+1}},
	\end{align*}
	\noindent then for any $\bw \in \Omega=\X \times \Y \times \Lambda$, we have
	\begin{equation*}
	P_1+(\bw-\bw_{k+1})^TG(\bw_{k+1}-\bw_k) \geq (1-\alpha-\gamma)\beta\|\br_{k+1}\|^2+(1-\alpha)\beta\langle \br_{k+1}, B(\by_k-\by_{k+1})\rangle+\langle \bw_{k+1}-\bw,F(\bw_{k+1})\rangle.
	\end{equation*}
\end{lem}

\noindent Next, we will simplify Lemma 1 based on different intervals of $\gamma$ in following three Lemmas. Note, we always need $\alpha\in[0,1)$.

\begin{lem}
	Suppose $\gamma\in(0,1)$, then there exists a constant $c_1$ (depending on $\alpha\ \&\ \gamma$) in (0,1), such that for any $\bw\in\Omega$, we have
	\begin{align*}
	2P_1+\|\bw_k-\bw\|_G^2-\|\bw_{k+1}-\bw\|_G^2\geq c_1\|\bw_{k}-\bw_{k+1}\|_G^2+2\langle\bw_{k+1}-\bw,F(\bw_{k+1})\rangle.
	\end{align*}
\end{lem}

\begin{lem}
	Suppose $\gamma=1$, then there exists a constant $c_2$ in (0,1), such that for any $\bw\in\Omega$, we have
	\begin{align*}
	2P_1+(\|\bw_k-\bw\|_G^2+c_2\|\by_k&-\by_{k-1}\|_T^2)-(\|\bw_{k+1}-\bw\|_G^2+c_2\|\by_{k+1}-\by_{k}\|_T^2)\\
	&\geq c_2\|\bw_k-\bw_{k+1}\|_G^2+2\langle\bw_{k+1}-\bw,F(\bw_{k+1})\rangle
	\end{align*}
\end{lem}

\begin{lem}
	Suppose $\gamma \in (1,\frac{1-\alpha+\sqrt{(1+\alpha)^2+4(1-\alpha^2)}}{2})$, then there exist the same constant $c_2$ as in Lemma 2 and another two constants $c_3$ and $\tau$ in (0,1), such that for any $\bw\in\Omega$, we have
	\begin{align*}
	2P_1+(\|\bw_k-\bw\|_G^2+c_2\|\by_k&-\by_{k-1}\|_T^2+c_3\beta\|\br_k\|^2)\\
	&-(\|\bw_{k+1}-\bw\|_G^2+c_2\|\by_{k+1}-\by_{k}\|_T^2+c_3\beta\|\br_{k+1}\|^2)\\
	&\geq\tau\|\bw_k-\bw_{k+1}\|_G^2+2\langle\bw_{k+1}-\bw,F(\bw_{k+1})\rangle
	\end{align*}
\end{lem}
\noindent Based on all the above lemmas, we can derive the convergence rate of the averaged iterates $\bar{\bu}$ by the following two theorems:

\begin{thm}[Main Theorem]
	Suppose $\alpha\in[0,1)$ and $\gamma\in(0,\frac{1-\alpha+\sqrt{(1+\alpha)^2+4(1-\alpha^2)}}{2})$, if $\theta_1, \theta_2$ are convex and $\eta_k = Ck^{-p}$ with $p\in(0,1)$ and a positive constant $C$, then under Assumptions 1 and 2, the averaged iterates $\bar{\bw}_t=(\bar{\bx}_t, \bar{\by}_t, \bar{\blambda}_t)^T$ generated by Algorithm~\ref{alg:5} satisfy: there exists a constant $D>0$ such that  $\forall\epsilon>0$, $\exists\  t$ (depending on $\epsilon$) large enough
	\begin{align*}
	E[\theta(\bar{\bu}_t)-\theta(\bu^*) + \langle\bar{\bw}_t-\bw^*, F(\bw^*)\rangle]
	\leq \frac{D}{2t} + \frac{D_X^2}{2Ct^{1-p}} + \frac{N^2C}{2(1-p)t^{p}}+\epsilon.
	\end{align*}
	Moreover, if we write by constraint, setting $p=\frac{1}{2}$, then $\forall \rho>0$ we have
	\begin{displaymath}
	E[ \theta(\bar{\bu}_t) - \theta(\bu^*) + \rho\|A\bar{\bx}_t+B\bar{\by}_t-\bb\|_2 ] = O(\frac{1}{\sqrt{t}}).
	\end{displaymath}
\end{thm}
\noindent Based on the above theorem, we can see the convergence rate for our algorithm is $O(1/\sqrt{t})$ in the ergodic sense.

\begin{thm}[Strong Convexity Case]
	If $\theta_1$ is a $\mu$-strongly convex function, setting $\eta_k = \frac{1}{k\mu}$, under Assumption 1 and 2, we have $\forall\epsilon>0$, $\exists\  t$ large enough such that
	\begin{align*}
	E[\theta(\bar{\bu}_t)-\theta(\bu^*) + \langle\bar{\bw}_t-\bw^*, F(\bw^*)\rangle]
	\leq  \frac{D+\mu D^2_X}{2t}+\frac{N^2\log t}{2\mu t}+\epsilon.
	\end{align*}
	Similarly, in another way, $\forall \rho>0$, we have
	\begin{displaymath}
	E[ \theta(\bar{\bu}_t) - \theta(\bu^*) + \rho\|A\bar{\bx}_t+B\bar{\by}_t-\bb\|_2 ] = O(\frac{\log t}{t}).
	\end{displaymath}
\end{thm}
\noindent Based on this theorem, we can see our proposed algorithm has $O(\log(t)/t)$ convergence rate if $\theta_1$ is strongly convex function.\\
Note that the convergence rate and the assumptions are exactly the same as the Stochastic ADMM~\cite{OH13,SA14a},
but we observe a faster contraction speed in practice due to the new update rule for $\blambda$. Comparing
to~\cite{SA14a}, the convergence rate is also the same under the mild condition that $\theta_1$ is convex
but may not be strongly convex.

\section{Numerical Experiments}
\label{sec:4}
In this section, we apply our Stochastic SPB-SPRSM algorithm to some models in statistical learning and compare the iteration efficiency with stochastic ADMM-typed algorithms. We focus on Lasso, Group Lasso, and Sparse logistic regression problems, derive the update rule for each model, and conduct experiments on both simulated and real datasets.

{\bf Competing Methods: }
We include the following ADMM-typed algorithms for solving the stochastic optimization problem~\eqref{eq:1-1} in
our comparison:

\begin{compactenum}
  \item Sto-SPB-PRSM: Our proposed method (Algorithm~\ref{alg:6}).
  \item SADMM: The algorithm proposed in~\cite{OH13} (Algorithm~\ref{alg:5}).
  \item OpSADMM: The algorithm proposed in~\cite{SA14a}.
  \item BatchADMM: The original ADMM algorithm.
  \end{compactenum}
  Note that all the above methods focus on solving the stochastic version of the problem~\eqref{eq:1-1},
  and they have the same time complexity for each update, so we present the objective function versus number
  of iterations in all the experiments. The algorithm proposed in~\cite{LWZ14a} is an incremental algorithm
  that will use previous samples, so cannot be compared directly with the above four algorithms.

{\bf Parameter Settings: }
As talked about in SPRSM \cite{HL14}, the underdetermined relaxation factor $\alpha$ is easily determined. In fact, empirically, $\alpha\in [0.8,0.9]$. Here, for simplicity, we take $\alpha=\gamma =0.9$ in all simulations and examples and all simulation parameters are set similarly as in \cite{HL14}. Also, all the three algorithms need $\beta$ (see Algorithm~\ref{alg:4} and \ref{alg:5}) and we follow~\cite{HL14,SA14a} to set $\beta=1$. To show the flexibility of our proposed algorithm, we set $S=I_m$ and $T=0$, which means we can add the proximal term in subproblem of $\bx$.

\subsection{Lasso}\label{sec:4-1}
The Lasso model can be formulated as:
\begin{equation}
	\min_\bx \big\{\frac{1}{2} \|D\bx-\br\|_2^2+\mu \|\bx\|_1\big\}
  \label{eq:lasso}
\end{equation}
where $\bx\in\R^d$ is the parameters,  $\br\in\mathfrak{R}^n$ is the response vector, $D\in\mathfrak{R}^{n\times d}$ is the design matrix with $n$ sample points and $d$ features, $\mu >0$ is the regularization parameter, $\|\cdot\|_1$ is the $\ell_1$-norm. For estimating a sparse parameter vector, we focus on the high dimensional problems ($n<d$). \\

To generate synthetic data, we draw each entry of the design matrix $D$ from $N(0,1)$, and generate
the underlying sparse $d$-dimensional parameter vector $\bar{\bx}$ with $100$ nonzero entries from $N(0,1)$.
The noise vector $\bepsilon$ is drawn from $N(0, 10^{-3}I)$, and the response vector $\br=D\bar{\bx}+\bepsilon$.
The regularization parameter is set as $\mu=0.1\|D^T\br\|_{\infty}$, where we found the recovered
entries have the similar number of nonzeroes with the underlying matrix $\bar{\bx}$.
Using this approach, we generate two synthetic data for the Lasso problem, as shown in Table~\ref{tab:1}.
\begin{table}[h]
	\centering
	\caption{Lasso Simulations}
	\label{tab:1}
	\begin{tabular}{|c|c|c|c|}
		\hline
		Type & d & n & $\eta_0$ \\
		\hline
		Simulation 1 & 400 & 200 & 1e-5 \\
		\hline
		Simulation 2 & 1000 & 500 & 1e-6 \\
		\hline
	\end{tabular}
\end{table}

Now, we show how to use Sto-SPB-SPRSM to solve the Lasso problem~\eqref{eq:lasso}. The problem can be rewritten as
\begin{equation*}
  \min_{\bx, \by} \frac{1}{2} \|D\bx-\br\|_2^2 + \mu \|\by\|_1 \ \text{ s.t. } \ \bx-\by=0,
\end{equation*}
which is equivalent to the stochastic optimization problem~\eqref{eq:1-1} with
$\theta_1(\bx, \xi) = \frac{1}{2} (\bd_{\xi}^T \bx - r_{\xi})^2$ and $\theta_2(\by) = \mu \|\by\|_1$,
where $\xi$ uniformly distributed in $\{1,2,...n\}$ and $\bd_{\xi}$ is the transpose of the $\xi$-th row of $D$.
When applying Algorithm~\ref{alg:6}, the update rules can be derived as shown in Algorithm~\ref{alg:7},
where the update rule for $\by$ is written by
\begin{equation*}
  \by_{k+1} \leftarrow S_{\frac{\mu}{\beta}}(\bx_{k+1} - \blambda_{k+1/2}/\beta),
\end{equation*}
where $S_{a}(\bx)$ is the soft-thresholding operator defined by
\begin{equation*}
  S_{a}(\bx) = \begin{cases}
    x_i-a, \ &\text{ if } x_i > a, \\
    0, \ &\text{ if } |x_i| \leq a, \ \ \forall i. \\
    x_i+a, \ &\text{ if } x_i < -a
  \end{cases}
\end{equation*}
\begin{algorithm}[tb]
	\caption{Stochastic SPB-SPRSM (Lasso)}
	\label{alg:7}
	\begin{algorithmic}
		\STATE Initialize $\bx_0$, $\by_0$ and $\blambda_0$ to be $\b0$.
		\FOR{$k=0, 1, 2,\ldots$}
    \STATE Sample $i$ from $\{1, \cdots, n\}$
    \STATE $\bx_{k+1} \leftarrow   \frac{  (r_{i}-\bd_{i}^T \bx_k)\bd_{i}+\blambda_k+\beta \by_k + (1+1/\eta_{k+1})\bx_k  }{\beta+1+1/\eta_{k+1}}$.
		\STATE $\blambda_{k+1/2}\leftarrow   \blambda_k - \alpha\beta(\bx_{k+1}-\by_{k})$.
        \STATE $\by_{k+1}\leftarrow    S_{\frac{\mu}{\beta}}(\bx_{k+1}-\blambda_{k+1/2}/\beta)$.
		\STATE $\blambda _{k+1}\leftarrow   \blambda_{k+1/2} - \gamma\beta(\bx_{k+1}-\by_{k+1})$.
		\ENDFOR
	\end{algorithmic}
\end{algorithm}

\begin{figure*}[tb]
  \centering
  \begin{tabular}{cc}
    \subfloat[Lasso Simulation 1\label{fig:1}]{\includegraphics[width=0.4\textwidth]{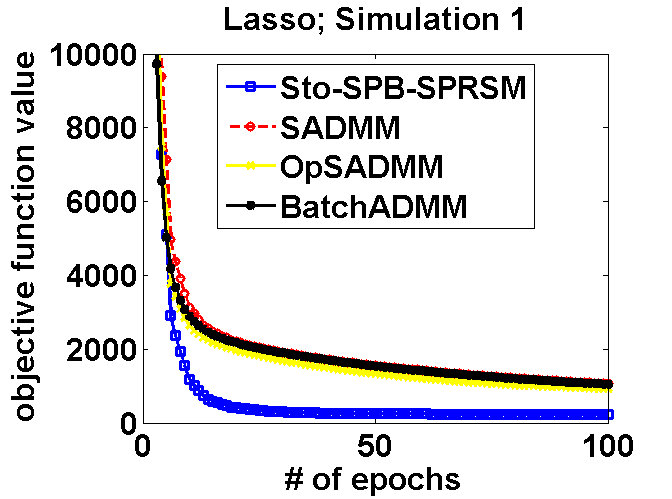}}
    &
    \subfloat[Lasso Simulation 2\label{fig:2}]{\includegraphics[width=0.4\textwidth]{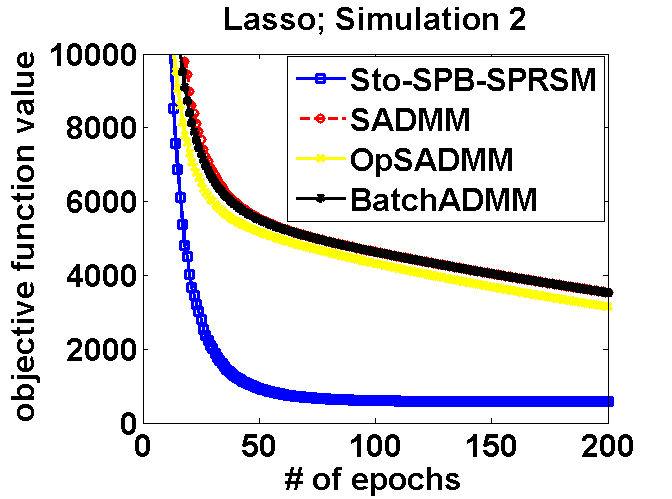}} \\
    \subfloat[Group Lasso\label{fig:3}]{\includegraphics[width=0.4\textwidth]{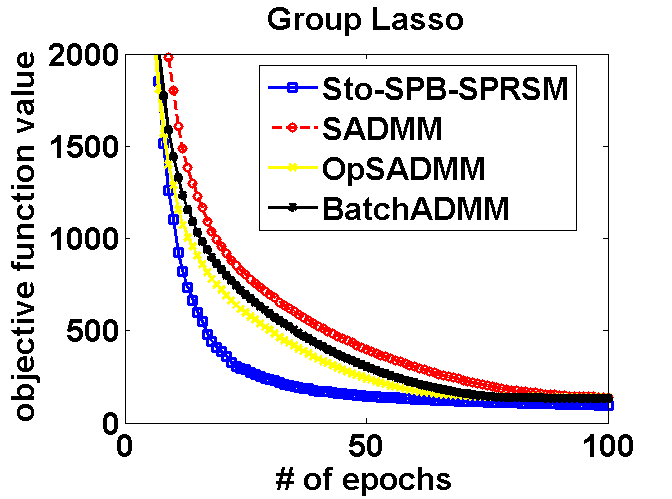}}
    &
    \subfloat[Logistic Regression\label{fig:4}]{\includegraphics[width=0.4\textwidth]{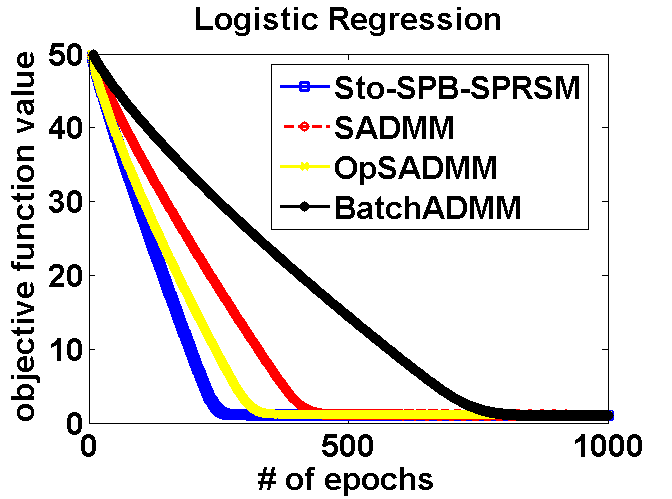}}
  \end{tabular}
  \caption{
Comparison with other stochastic optimization algorithms on synthetic datasets.
  Our proposed algorithm (Sto-SPB-SPRSM) converges faster than other methods. }
\end{figure*}

The experimental results are shown in Fig\ref{fig:1} and Fig\ref{fig:2}. We can observed that in both settings our proposed algorithm is faster than existing algorithms.

\subsection{Group Lasso}
The Group Lasso model can be formulated as:
\begin{equation*}
  \min_{\bx} \big\{\frac{1}{2} \|D\bx-\br\|_2^2 + \mu \sum_{i=1}^N \|\bx_i\|_2 \big\}
\end{equation*}
where $N$ is the number of disjointed groups and $\bx_i\in\mathfrak{R}^{d_i}$ is the $d_i$-dimensional parameter vector of the $i$-th group. All other settings are the same as Lasso model.

We generate the synthetic dataset by the following way: We set $n=200$ and generate $N=10$ blocks with size $d_i$ uniformly distributed between $1$ and $50$. $d=\sum_{i=1}^{N}d_i$. For the parameter $\bx_i$, $5\%$ of entries are drawn from the standard normal distribution with the rest set to be zero. We set $\mu=0.1\max\{\|\bd_1^T\br\|_\infty, \cdots, \|\bd_N^T \br\|_{\infty}\}$. For the design matrix $D$ and response vector $r$, we use the same method as (\ref{sec:4-1}).

Next we derive the update rule of stochastic SPB-SPRSM for solving the Group Lasso problem. Here,
$\theta _1 (\bx,\xi )$ is still defined as $\theta _1 (\bx,\xi )=\frac{1}{2} (\bd_{\xi}^T \bx - r_{\xi})^2$ where $d_{\xi}$ is the transpose of the $\xi$-th row of design matrix $D$ and $r_i$ is the $i$-th entry of response vector $r$. The update rules can be derived as shown in Algorithm \ref{alg:8},
where the update rule for $\by$ is again the soft-thresholding operator but for $L_2$ norm (block soft thresholding), which means $S_{k}(\bx)$ here is defined by
$(S_k(\bx))_i=(1-k/\|x_i\|_2)_+\times x_i, i=1,2,...d$.
Note that in Algorithm~\ref{alg:8}, we only need to update $\by$ for the current group at each iteration.

The experimental results are show in Fig~\ref{fig:3}. We can observed that in sparse group lasso, our proposed algorithm still converges faster than other existing algorithms.

\begin{algorithm}[tb]
	\caption{Stochastic SPB-SPRSM (Group Lasso)}
	\label{alg:8}
	\begin{algorithmic}
		\STATE Initialize $\bx_0$, $\by_0$ and set $\blambda_0=0$.
		\FOR{$k=0, 1, 2,\ldots$}
    \STATE Sample $i$ from $\{1, \cdots, n\}$
    \STATE $\bx_{k+1} \leftarrow   \frac{  (r_{i}-\bd_{i}^T \bx_k)\bd_{i}+\blambda_k+\beta \by_k + (1+1/\eta_{k+1})\bx_k  }{\beta+1+1/\eta_{k+1}}$.
		\STATE $\blambda_{k+1/2}\leftarrow   \blambda_k - \alpha\beta(\bx_{k+1}-\by_{k})$.
        \STATE $\by_{i,k+1}\leftarrow    S_{\frac{\mu}{\beta}}(\bx_{i,k+1}-\blambda_{i,k+1/2}/\beta),i=1,2,...,N$.
		\STATE $\blambda_{k+1}\leftarrow   \blambda_{k+1/2} - \gamma\beta(\bx_{k+1}-\by_{k+1})$.
		\ENDFOR
	\end{algorithmic}
\end{algorithm}

\subsection{Sparse Logistic Regression}
The sparse logistic regression model can be written as:
\begin{equation*}
  \min_{\bx} \big\{ \frac{1}{n}\sum_{i=1}^n \log(1+\exp(-r_i (\bd_i^T \bx+x_0)))+ \mu \|\bx\|_1 \big\},
\end{equation*}
where $n$ is the number of sample points; $\bd_i$ is the $i$th row of the design matrix. Moreover, $\bx\in\mathfrak{R}^d$ is the $d$-dimensional parameter vector; $r_i\in\{1,-1\} (i=1,2,\ldots ,n)$ are the $i$th response value.
To generate the synthetic dataset, as the method in (\ref{sec:4-1}), we draw each entry of the normalized $n\times d$ design matrix $D$ from $N(0,1)$, a sparse $d$-dimensional parameter vector $\bx$ with $100$ nonzero entries from $N(0,1)$, the noise vector $\bepsilon$ from $N(0,10^{-3} I)$, the response vector is $r=\text{sign}(D\bx+\bepsilon )$. For simplicity, we set $\mu =1$.

Similar to the previous two cases, we can transform the sparse logistic regression problem into~\eqref{eq:1-1} by setting
$\theta_1(\bx, \xi)=\log(1+\exp(-r_{\xi} \bd_{\xi}^T \bx))$ and $\theta_2(\by)=\mu\|\by\|_1$.
We can then derive the update rule, as shown in Algorithm~\ref{alg:9}. Note that the step for updating $\by$ is the same as
the Lasso problem. The simulation results are shown in Fig.\ref{fig:4}. We can observed that in sparse logistic regression model, our proposed algorithm still converges faster than other algorithms.

\begin{algorithm}[tb]
	\caption{Stochastic SPB-SPRSM (Logistic Regression)}
	\label{alg:9}
	\begin{algorithmic}
		\STATE Initialize $x_0$, $y_0$ and set $\lambda _0=0$.
		\FOR{$k=0, 1, 2,\ldots$}
    \STATE Sample $i$ from $\{1, \cdots, n\}$
    \STATE $\bx_{k+1}\leftarrow     \frac{ (r_{i}\bd_i / (1+\exp(r_{i}(\bd_i^T \bx_k+x_0)))+\blambda_k+\beta \by_k+(1+1/\eta_{k+1})\bx_k}{\beta+1+1/\eta_{k+1}}$.
		\STATE $\blambda_{k+1/2}\leftarrow     \blambda_k-\alpha\beta(\bx_{k+1}-\by_k)$.
		\STATE $\by_{k+1}\leftarrow      S_{\frac{\mu}{\beta}}(\bx_{k+1}-\blambda_{k+1/2}/\beta)$.
		\STATE $\blambda _{k+1}\leftarrow     \blambda_{k+1/2}-\gamma\beta(\bx_{k+1}-\by_{k+1})$.
		\ENDFOR
	\end{algorithmic}
\end{algorithm}

\begin{figure*}[tb]
  \centering
  \begin{tabular}{ccc}
    \subfloat[bodyfat, $\mu=10$ \label{fig:5}]{\includegraphics[width=0.32\textwidth]{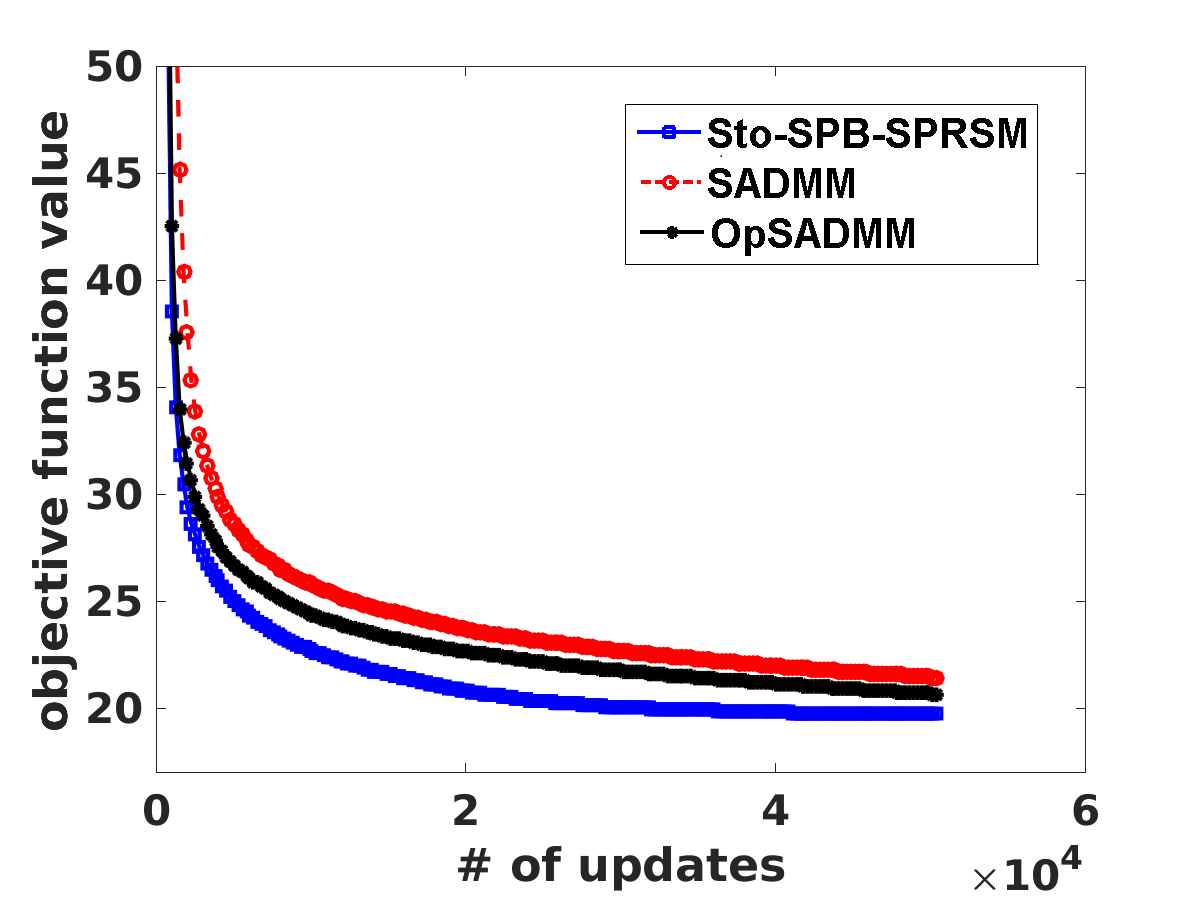}}
    &
    \subfloat[a9a, $\mu=10$\label{fig:6}]{\includegraphics[width=0.32\textwidth]{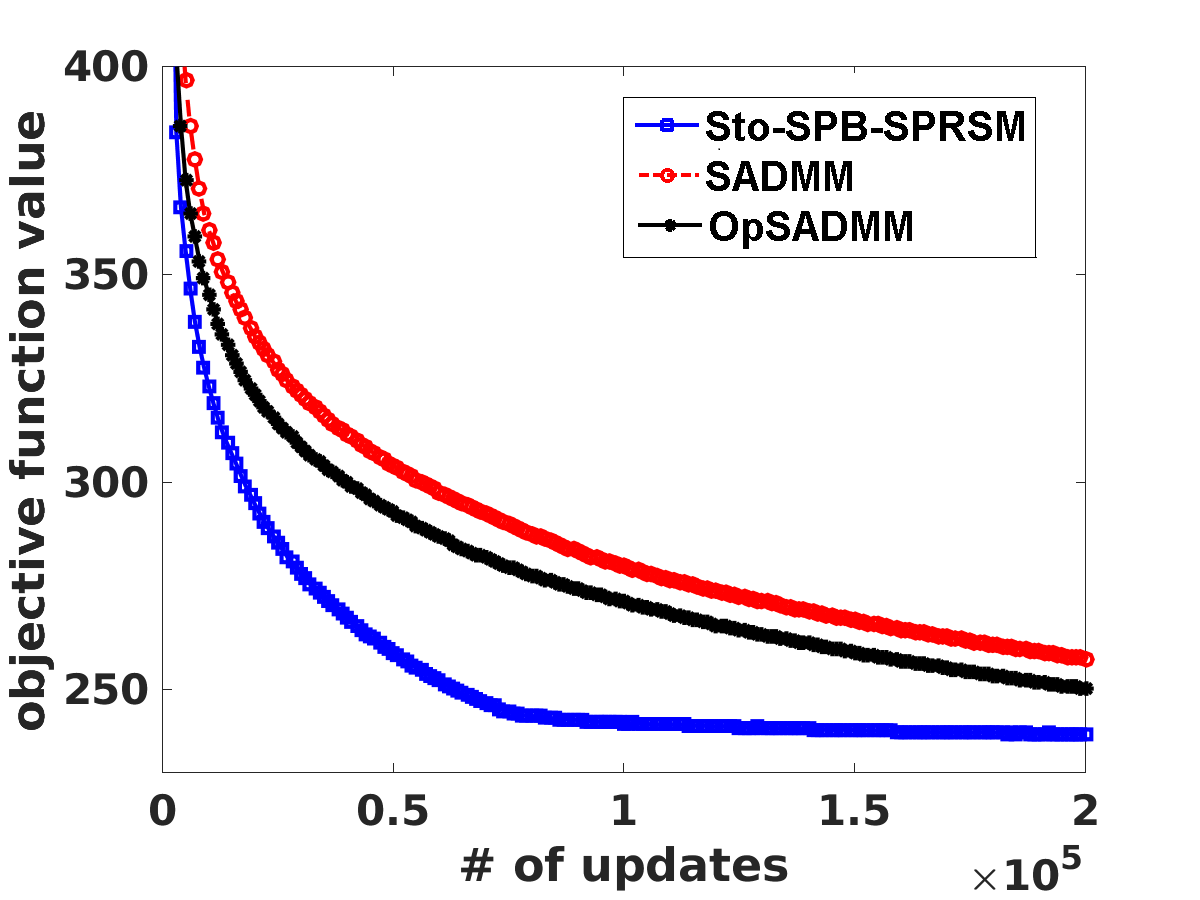}}
    &
    \subfloat[E2006, $\mu=10$ \label{fig:7}]{\includegraphics[width=0.32\textwidth]{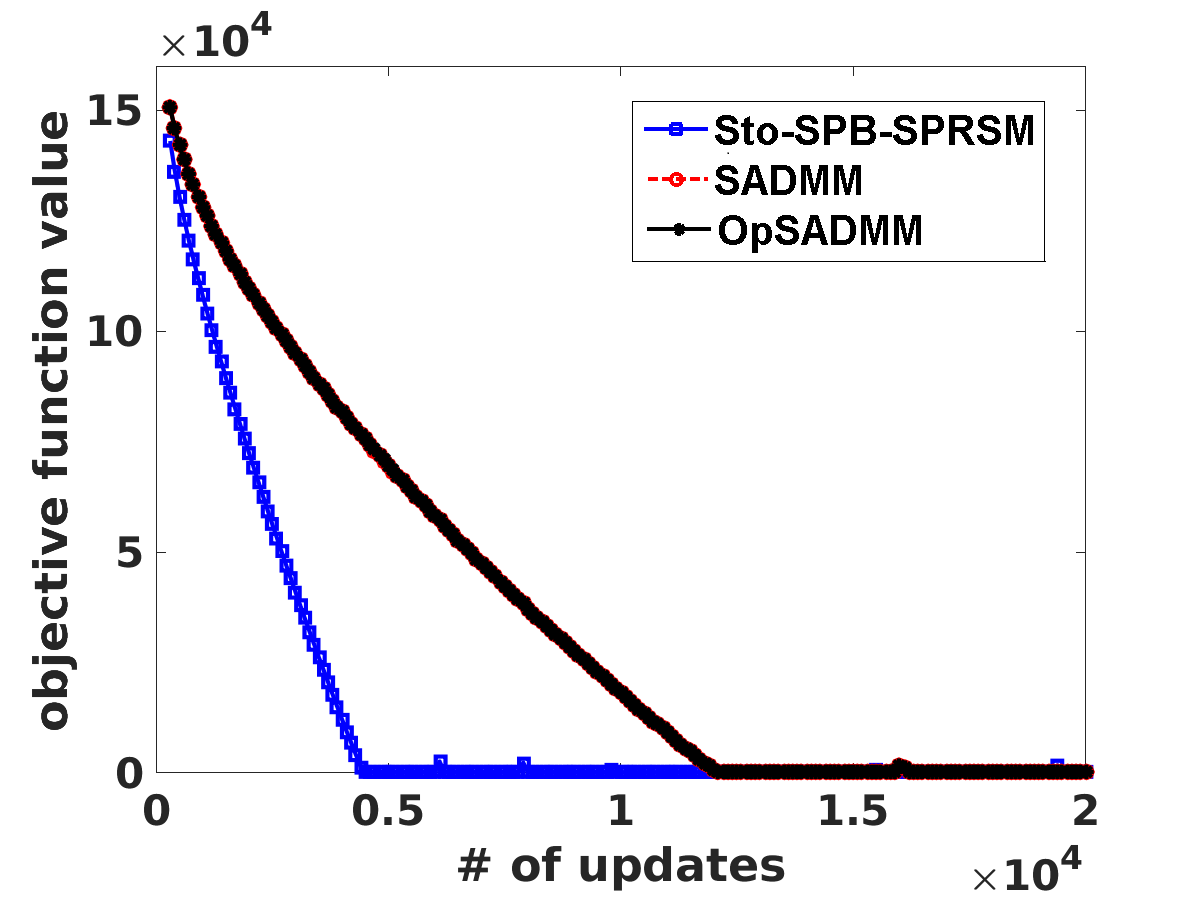}} \\
    \subfloat[bodyfat, $\mu=1$ \label{fig:5}]{\includegraphics[width=0.32\textwidth]{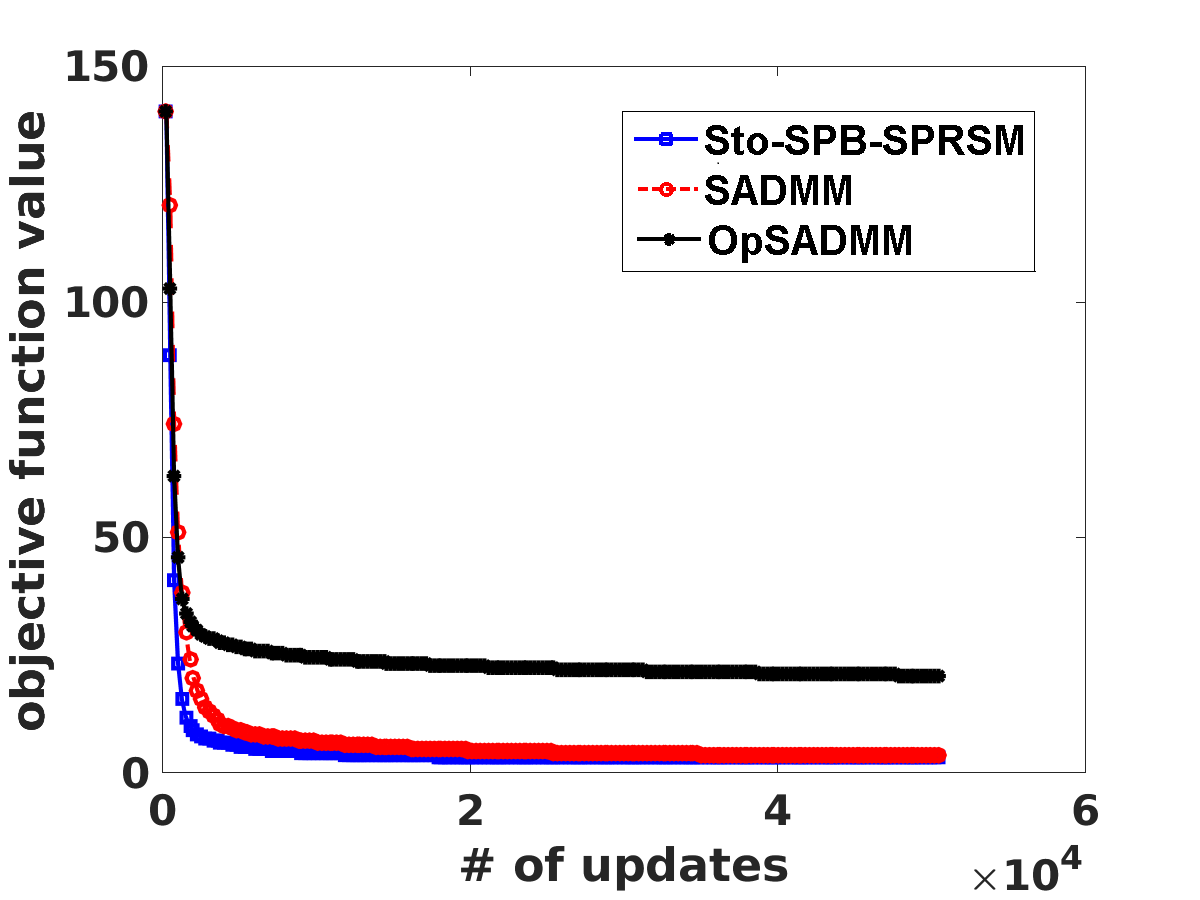}}
    &
    \subfloat[a9a, $\mu=1$ \label{fig:6}]{\includegraphics[width=0.32\textwidth]{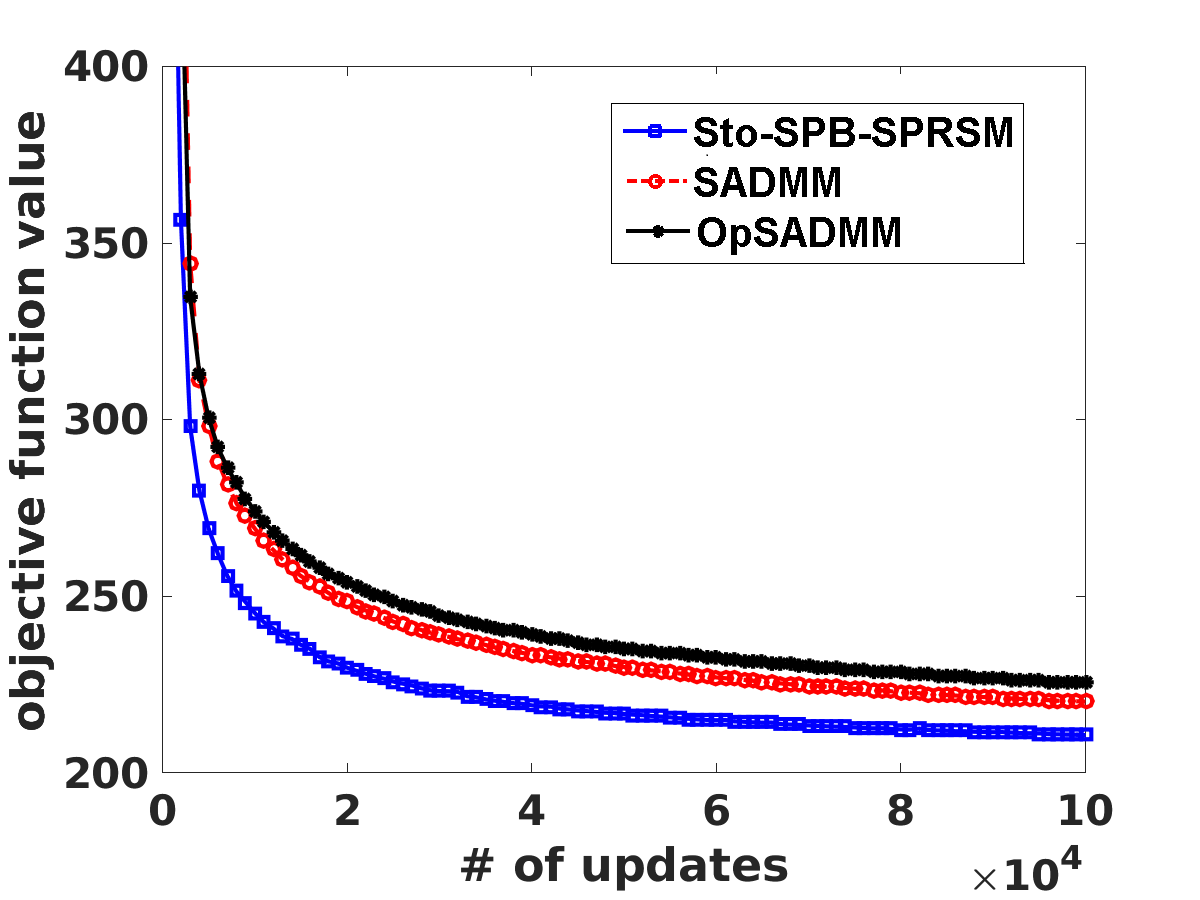}}
    &
    \subfloat[E2006, $\mu=1$ \label{fig:7}]{\includegraphics[width=0.32\textwidth]{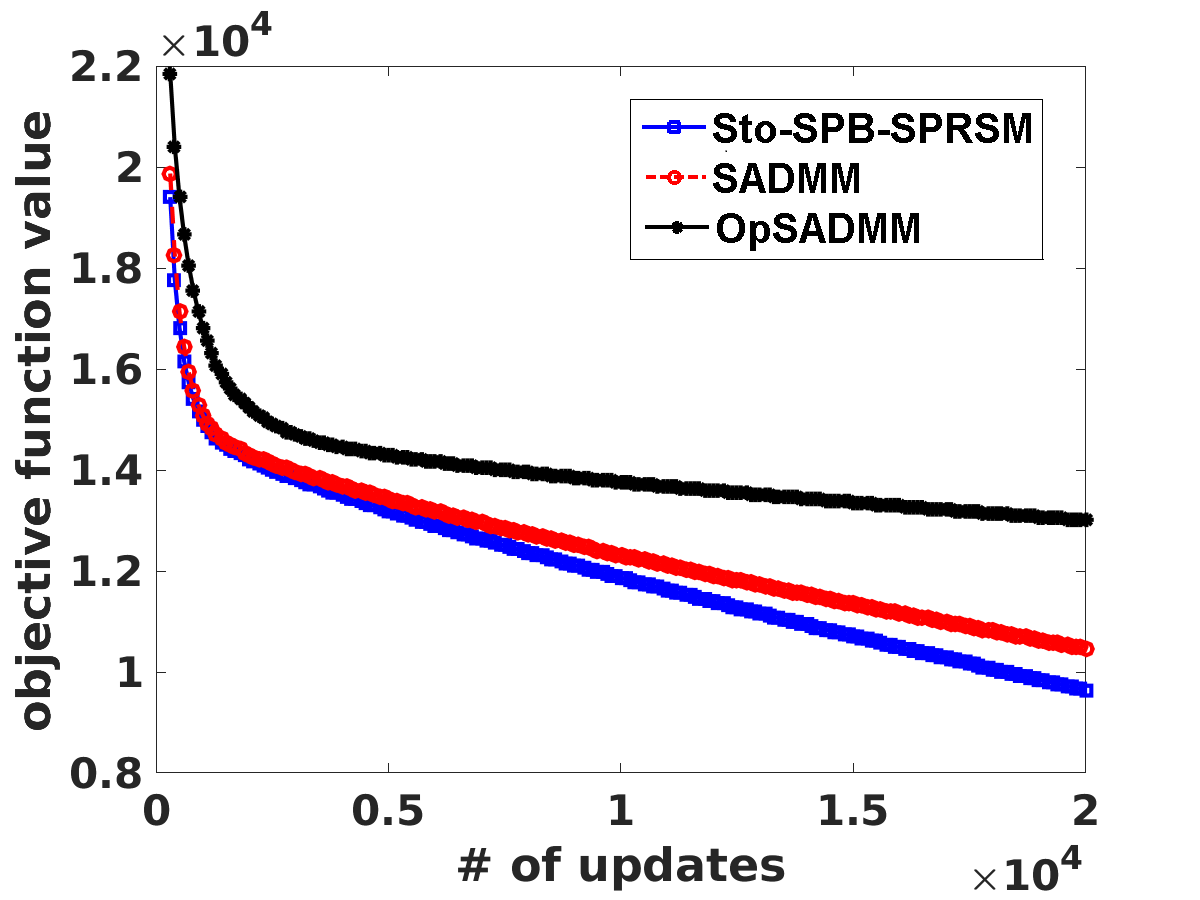}}
  \end{tabular}
  \caption{Comparison with other stochastic optimization algorithms on real datasets.
  The $x$-axis is number of updates (each using only one training sample).
  Note that
  SADMM and OpSADMM almost overlap on the E2006 dataset. We observe that
  our proposed algorithm (Sto-SPB-SPRSM) converges faster than other methods, especially when the solution is sparse
  ($\mu=10$).
  }
\end{figure*}

\subsection{Comparisons on Real Datasets}

Finally, we compare our proposed algorithm with existing ADMM-typed algorithms
on real datasets. We test the convergence speed for solving the Lasso problem,
and we consider the following three datasets in Table~\ref{tab:2}. For simplicity,
we test all the algorithms with $\mu=10, 1$ for all the three datasets.
Note that when $\mu=10$, the solution is sparse, while the solution will be dense when $\mu=1$
for all the datasets.
For the step size, we set $\eta_{\text{init}}=10^{-6}$ for all the methods.
Note that the BatchADMM converges much slower than other methods, so we ignore the comparison
here.
The results are shown in Figure~\ref{fig:5}, \ref{fig:6} and \ref{fig:7}. We can clearly see that
our proposed SSPRSM algorithm converges faster than other methods, especially when $\mu=10$ (which means
the solution is sparse).

\begin{table}[h]
	\centering
	\caption{Real Datasets \label{tab:2}
  }
	\begin{tabular}{|c|c|c|c|}
		\hline
		 & $d$ & $n$ & $\gamma$ \\
		\hline
	  Bodyfat & 14 & 252 & 1, 10 \\
		\hline
	a9a & 123 & 32,561 & 1, 10 \\
		\hline
    E2006 & 150,360 & 16,087 & 1, 10  \\
    \hline
	\end{tabular}
\end{table}

\section{Summary and future work}
\label{sec:summary}
In this paper, we have proposed another variant of SPRSM: Stochastic SPB-SPRSM. Using approximated augmented Lagrange function, our proposed algorithm can be applied to a general class of stochastic optimization problem with linear constraints, where the proximal function may not be easily computable.
Moreover, in our proposed algorithm, each iteration only requires one or a small subset of samples, which is suitable for large-scale
machine learning problems with large number of samples.
Furthermore, we proved the $O(1/\sqrt{t})$ convergence rate for convex functions, and $O(\log(t)/t)$ convergence rate for
strongly convex function. Experimental results show that our proposed algorithm is much faster than existing algorithms published in the past few years on real datasets.

Based on the main task: solving the subproblem, $x$-optimization problem, of Stochastic SPB-SPRSM in more general machine learning model, such as Graph-Guided Support Vector Machine, we may consider a more general stochastic algorithm where we can not only strength the convergence rate but also make subproblem easy to solve. Moreover, our splitting problem is only about two separable functions. So, applying our algorithm to a more general splitting problem, where we may have $n$ separable functions, is also our research topic. 
\newpage
\bibliography{reference.bib}
\bibliographystyle{plain}
\newpage
\section{Appendix}

We first summary the iteration scheme of Stochastic SPB-SPRSM algorithm. We define the first-order approximated augmented Lagrangian function as follows:
\begin{align*}
\hat{L}_{\beta ,k}(\bx,\by,\blambda )= \theta_1(\bx_k) &+ \langle \theta_1'(\bx_k, \xi_{k+1}), \bx\rangle +\theta_2(\by) - \langle \blambda , A\bx+B\by-\bb\rangle\\
&+\frac{\beta}{2}\|A\bx+B\by-\bb\|^2+\frac{\|\bx-\bx_k\|^2}{2\eta_{k+1}}
\end{align*}
The Stochastic SPB-SPRSM is equivalent to minimize the $\hat{L}_{\beta,k}(\bx,\by,\blambda)$. We have its update scheme:\\
\begin{equation*}
\left\{
\begin{aligned}
&\bx_{k+1}=\arg\min_{\bx\in \X} \left\{\hat{L}_{\beta,k}(\bx,\by_k,\blambda_k)+\frac{\|\bx-\bx_k\|^2_S}{2}\right\}\\
&\blambda_{k+1/2} = \blambda _k-\alpha\beta(A\bx_{k+1}+B\by_k-\bb)\\
&\by_{k+1}=\arg\min_{\by\in \Y} \left\{\hat{L}_{\beta,k}(\bx_{k+1},\by,\blambda_{k+1/2})+\frac{\|\by-\by_k\|^2_T}{2}\right\}\\
&\blambda_{k+1} = \blambda_{k+1/2}-\gamma\beta(A\bx_{k+1}+B\by_{k+1}-\bb)
\end{aligned}
\right.
\end{equation*}
Thus, plug in $\hat{L}_{\beta,k}(\bx,\by,\blambda)$ and we get the final iteration scheme:\\
\begin{equation*}
\left\{
\begin{aligned}
&\bx_{k+1}=\arg\min_{\bx\in \X} \left\{\langle \theta_1'(\bx_k,\xi_{k+1}),x\rangle-\blambda_k^TA\bx+\frac{\beta}{2}\|A\bx+B\by_k-\bb\|^2+\frac{\|\bx-\bx_k\|^2}{2\eta_{k+1}}+\frac{\|\bx-\bx_k\|^2_S}{2}\right\}\\
&\blambda_{k+1/2} = \blambda _k-\alpha\beta(A\bx_{k+1}+B\by_k-\bb)\\
&\by_{k+1}=\arg\min_{\by\in \Y}\left\{ \theta_2(\by)-\blambda_{k+1/2}^TB\by+\frac{\beta}{2}\|A\bx_{k+1}+B\by-\bb\|^2+\frac{\|\by-\by_k\|^2_T}{2}\right\}\\
&\blambda_{k+1} = \blambda_{k+1/2}-\gamma\beta(A\bx_{k+1}+B\by_{k+1}-\bb)
\end{aligned}
\right.
\end{equation*}

\subsection*{7.1 Proof of Lemma 1}
Applying the optimality condition for the iteration of $\bx$, we have $\forall \bx \in \X$
\begin{align*}
\langle \bx-\bx_{k+1},  &\theta_1'(\bx_k,\xi_{k+1}) \rangle + \langle \bx-\bx_{k+1},\frac{1}{\eta_{k+1}}(\bx_{k+1}-\bx_k)\rangle\\ 
&+\langle \bx-\bx_{k+1}, S(\bx_{k+1}-\bx_k)
-A^T\blambda_k+\beta A^T(A\bx_{k+1}+B\by_k-b)\rangle\geq 0.
\end{align*}
We will simplify this inequality term by term. For the first term on the left hand side, we have
\begin{align*}
\langle \bx-\bx_{k+1}, \theta_1'(\bx_k,\xi_{k+1}) \rangle
&=\langle \bx-\bx_{k},  \theta_1'(\bx_k) \rangle+\langle \bx-\bx_{k},  \theta_1'(\bx_k,\xi_{k+1})-\theta_1'(\bx_k) \rangle+\langle \bx_k-\bx_{k+1},  \theta_1'(\bx_k,\xi_{k+1}) \rangle\\
&=\langle \bx-\bx_{k},  \theta_1'(\bx_k) \rangle+\langle \bx-\bx_{k},  \delta_{k+1} \rangle+\langle \bx_k-\bx_{k+1},  \theta_1'(\bx_k,\xi_{k+1}) \rangle\\
&\leq \theta_1(\bx)-\theta_1(\bx_k)+\langle \bx-\bx_k,\delta_{k+1}\rangle+\frac{1}{2\eta_{k+1}}\|\bx_k-\bx_{k+1}\|^2+\frac{\eta_{k+1}}{2}\|\theta_1'(\bx_k,\xi_{k+1})\|^2. \tag{1}
\end{align*}
The second equality is because of the definition of $\delta_{k+1}$ and the last inequality is Cauchy-Schwarz inequality.\\
For the second term on the left hand side, we utilize the equaltiy
\begin{equation*}
\|\ba\|_G^2-\|\bb\|_G^2=\|\ba-\bb\|_G^2+2\bb^TG(\ba-\bb). \eqno(2)
\end{equation*}
Here, we set $\ba=\bx-\bx_k$ and $\bb=\bx-\bx_{k+1}$, then we get
\begin{equation*}
\langle \bx-\bx_{k+1},\frac{1}{\eta_{k+1}}(\bx_{k+1}-\bx_k)\rangle=\frac{1}{2\eta_{k+1}}(\|\bx-\bx_k\|^2-\|\bx-\bx_{k+1}\|^2-\|\bx_{k+1}-\bx_k\|^2). \eqno(3)
\end{equation*}
So, we combine (1) and (3) and get the following inequality of $\bx$,
\begin{align*}
&\theta_1(\bx)-\theta_1(\bx_k)+\langle \bx-\bx_k,\delta_{k+1}\rangle+\frac{\eta_{k+1}}{2}\|\theta_1'(\bx_k,\xi_{k+1})\|^2+\frac{1}{2\eta_{k+1}}(\|\bx-\bx_k\|^2-\|\bx-\bx_{k+1}\|^2)\\
&+\langle \bx-\bx_{k+1}, S(\bx_{k+1}-\bx_k)-A^T\blambda_k+\beta A^T(A\bx_{k+1}+B\by_k-\bb)\rangle \geq 0\ \ \ \forall \bx \in \X \tag{4}
\end{align*}
Applying the optimality condition for the iteration of $\by$, we directly get the inequality of $\by$, 
\begin{equation*}
\theta_2(\by)-\theta_2(\by_{k+1})+\langle \by-\by_{k+1},T(\by_{k+1}-\by_k)-B^T\blambda_{k+1/2}+\beta B^T(A\bx_{k+1}+B\by_{k+1}-\bb) \geq 0\ \ \ \ \ \  \forall \by \in \Y \eqno(5)
\end{equation*}
Based on the definition of $\br_k$, $\br(\bw)$, $\theta(\bu)$ and $F(\bw)$, we can further get the equality of $\blambda$:
\begin{equation*}
\begin{split}
\blambda_{k+1/2}&=\blambda_{k+1}+\gamma\beta \br_{k+1}\\
\blambda_{k}&=\blambda_{k+1/2}+\alpha\beta(A\bx_{k+1}+B\by_k-\bb)=\blambda_{k+1}+\gamma\beta\br_{k+1}+\alpha\beta\br_{k+1}+\alpha\beta B(\by_k-\by_{k+1})\\
&=\blambda_{k+1}+(\alpha+\gamma)\beta\br_{k+1}+\alpha\beta B(\by_k-\by_{k+1})
\end{split}
\end{equation*}
So, we have
\begin{equation*}
\br_{k+1}-\frac{\alpha}{\alpha+\gamma}B(\by_{k+1}-\by_k)+\frac{1}{(\alpha+\gamma)\beta}(\blambda_{k+1}-\blambda_k)=0. \eqno(6)
\end{equation*}
Then, based on the definition of $P_1$, we unify (4), (5) and (6)
\begin{align*}
P_1&+\langle\bw-\bw_{k+1}, \left( \begin{array}{ccc}
S(\bx_{k+1}-\bx_k) \\
T(\by_{k+1}-\by_k) \\
0  \end{array} \right)+\left( \begin{array}{ccc}
0 \\
\alpha\beta B^T\br_{k+1}+(1-\alpha)\beta B^T B(\by_{k+1}-\by_k) \\
-\frac{\alpha}{\alpha+\gamma}B(\by_{k+1}-\by_k)+\frac{1}{(\alpha+\gamma)\beta}(\blambda_{k+1}-\blambda_k)  \end{array} \right)\rangle\\
&+\langle \bw-\bw_{k+1},\left( \begin{array}{ccc}
A^T \\
B^T \\
0  \end{array} \right)((1-\alpha-\gamma)\beta\br_{k+1}+(1-\alpha)\beta B(\by_k-\by_{k+1}))\rangle \geq \langle \bw_{k+1}-\bw, F(\bw_{k+1})\rangle. \tag{7} 
\end{align*}
Next, we will further simplify the inequality (7). Denote the second term on the left hand side in (7) as $P_2$ and the third term as $P_3$, then we will deal with $P_2$ and $P_3$ respectively.\\
For $P_2$, using (6), we have
\begin{align*}
&\left( \begin{array}{ccc}
\alpha\beta B^T\br_{k+1}+(1-\alpha)\beta B^TB(\by_{k+1}-\by_k) \\
-\frac{\alpha}{\alpha+\gamma}B(\by_{k+1}-\by_k)+\frac{1}{(\alpha+\gamma)\beta}(\blambda_{k+1}-\blambda_k)  \end{array} \right) \\
&=\left( \begin{array}{ccc}
(1-\alpha)\beta B^TB(\by_{k+1}-\by_k)+\alpha\beta B^T(\frac{\alpha}{\alpha+\gamma}B(\by_{k+1}-\by_k)-\frac{1}{(\alpha+\gamma)\beta}(\blambda_{k+1}-\blambda_k)) \\
-\frac{\alpha}{\alpha+\gamma}B(\by_{k+1}-\by_k)+\frac{1}{(\alpha+\gamma)\beta}(\blambda_{k+1}-\blambda_k)  \end{array} \right)\\
&=\left( \begin{array}{ccc}
\frac{\alpha+\gamma-\alpha\gamma}{\alpha+\gamma}\beta B^TB(\by_{k+1}-\by_k)-\frac{\alpha}{\alpha+\gamma}(\blambda_{k+1}-\blambda_k) \\
-\frac{\alpha}{\alpha+\gamma}B(\by_{k+1}-\by_k)+\frac{1}{(\alpha+\gamma)\beta}(\blambda_{k+1}-\blambda_k)  \end{array} \right). \tag{8}
\end{align*}
Based on the definition of $H$, $P$ and $G$, we rewrite $P_2$ as
\begin{align*}
P_2=\langle\bw-\bw_{k+1}, \left( \begin{array}{ccc}
S(\bx_{k+1}-\bx_k) \\
T(\by_{k+1}-\by_k) \\
0  \end{array} \right)+\left( \begin{array}{ccc}
0 \\
H(\bv_{k+1}-\bv_k) \end{array} \right)\rangle=(\bw-\bw_{k+1})^TG(\bw_{k+1}-\bw_k).& \tag{9}
\end{align*}
For $P_3$, using $(A\  B\  0)(\bw_{k+1}-\bw)=\br_{k+1}-\br(\bw)=\br_{k+1}$, we have
\begin{align*}
-P_3&=\langle \br_{k+1}-\br(\bw),(1-\alpha-\gamma)\beta\br_{k+1}+(1-\alpha)\beta B(\by_k-\by_{k+1})\rangle\\
&=(1-\alpha-\gamma)\beta\|\br_{k+1}\|^2+(1-\alpha)\beta\langle \br_{k+1},B(\by_k-\by_{k+1})\rangle.& \tag{10}
\end{align*}
So, combine (9) and (10), we get final inequality in lemma 1
\begin{equation*}
P_1+(\bw-\bw_{k+1})^TG(\bw_{k+1}-\bw_k) \geq (1-\alpha-\gamma)\beta\|\br_{k+1}\|^2+(1-\alpha)\beta\langle \br_{k+1}, B(\by_k-\by_{k+1})\rangle+\langle \bw_{k+1}-\bw,F(\bw_{k+1})\rangle. \eqno(11)
\end{equation*}
\qed
\subsection*{7.2 Proof of Lemma 2}
Using the equality (2) again, we expand the second term on the left hand side
\begin{equation*}
(\bw-\bw_{k+1})^TG(\bw_{k+1}-\bw_k)=\frac{1}{2}(\|\bw_k-\bw\|_G^2-\|\bw_{k+1}-\bw\|_G^2-\|\bw_k-\bw_{k+1}\|_G^2), \eqno(12)
\end{equation*}
and based on the definition of $G$, we have
\begin{equation*}
\|\bw_k-\bw_{k+1}\|_G^2=\|\bu_k-\bu_{k+1}\|_P^2+\|\bv_k-\bv_{k+1}\|_H^2. \eqno(13)
\end{equation*}
Plug (12) and (13) into (11), we have
\begin{align*}
2P_1+\|\bw_k-\bw\|_G^2-\|\bw_{k+1}-\bw\|_G^2 \geq&\|\bu_k-\bu_{k+1}\|_P^2+\|\bv_k-\bv_{k+1}\|_H^2+2(1-\alpha-\gamma)\beta\|\br_{k+1}\|^2\\
&+2(1-\alpha)\beta\langle \br_{k+1}, B(\by_k-\by_{k+1})\rangle+2\langle \bw_{k+1}-\bw,F(\bw_{k+1})\rangle.
\end{align*}
Notice that
\begin{equation*}
\bv_k-\bv_{k+1}=\left( \begin{array}{ccc}
\by_k-\by_{k+1} \\
\blambda_k-\blambda_{k+1}  \end{array} \right)=\left( \begin{array}{ccc}
I_{n_2} & 0 \\
\alpha\beta B & (\alpha+\gamma)\beta I_m  \end{array} \right)\left( \begin{array}{ccc}
\by_k-\by_{k+1} \\
\br_{k+1}  \end{array} \right)=M\left( \begin{array}{ccc}
\by_k-\by_{k+1} \\
\br_{k+1}  \end{array} \right).
\end{equation*}
So, we have
\begin{align*}
\|\bv_k-\bv_{k+1}\|_H^2&=\left( \begin{array}{ccc}
\by_k-\by_{k+1} \\
\br_{k+1}  \end{array} \right)^TM^THM\left( \begin{array}{ccc}
\by_k-\by_{k+1} \\
\br_{k+1}  \end{array} \right)\\
&=\left( \begin{array}{ccc}
\by_k-\by_{k+1} \\
\br_{k+1}  \end{array} \right)^T\left( \begin{array}{ccc}
(1-\alpha)\beta B^TB & 0 \\
0 & (\alpha+\gamma)\beta I_m  \end{array} \right)\left( \begin{array}{ccc}
\by_k-\by_{k+1} \\
\br_{k+1}  \end{array} \right)\\
&=(1-\alpha)\beta\|B(\by_k-\by_{k+1})\|^2+(\alpha+\gamma)\beta\|\br_{k+1}\|^2. \tag{14}
\end{align*}
Plug this equality to above inequality, we have
\begin{align*}
2P_1+\|\bw_k-\bw\|_G^2-\|\bw_{k+1}-\bw\|_G^2 \geq&\|\bu_k-\bu_{k+1}\|_P^2+(1-\alpha)\beta\|B(\by_k-\by_{k+1})\|^2+(2-\alpha-\gamma)\beta\|\br_{k+1}\|^2\\
&+2(1-\alpha)\beta\langle \br_{k+1}, B(\by_k-\by_{k+1})\rangle+2\langle \bw_{k+1}-\bw,F(\bw_{k+1})\rangle. \tag{15}
\end{align*}
Moreover, we simplify (15) by using the definition of $K$ and $M$. We have
\begin{flalign*}
RHS&=\|\bu_k-\bu_{k+1}\|_P^2+2\langle\bw_{k+1}-\bw,F(\bw_{k+1})\rangle\\
&+\left( \begin{array}{ccc}
\by_k-\by_{k+1} \\
\br_{k+1}  \end{array} \right)^T\left( \begin{array}{ccc}
(1-\alpha)\beta B^TB & (1-\alpha)\beta B^T \\
(1-\alpha)\beta B & (2-\alpha-\gamma)\beta I_m  \end{array} \right)\left( \begin{array}{ccc}
\by_k-\by_{k+1} \\
\br_{k+1}  \end{array} \right)\\
&=\|\bu_k-\bu_{k+1}\|_P^2+2\langle\bw_{k+1}-\bw,F(\bw_{k+1})\rangle+ (\bv_k-\bv_{k+1})^TM^{-T}KM^{-1}(\bv_k-\bv_{k+1}).
\end{flalign*}
If $\gamma\in (0,1)$, then define $c_1=\frac{1-\sqrt{1-(\alpha+\gamma)(1-\gamma)}}{\alpha+\gamma} \in (0,1)$, we will have $K\succeq c_1M^THM$ where 
\begin{equation*}
M^THM=\left( \begin{array}{ccc}
(1-\alpha)\beta B^TB & 0 \\
0 & (\alpha+\gamma)\beta I_m  \end{array} \right).
\end{equation*}
So, we have final result
\begin{align*}
2P_1+\|\bw_k-\bw\|_G^2-\|\bw_{k+1}-\bw\|_G^2 &\geq\|\bu_k-\bu_{k+1}\|_P^2+2\langle\bw_{k+1}-\bw,F(\bw_{k+1})\rangle+c_1\|\bv_k-\bv_{k+1}\|_H^2\\
&\geq c_1\|\bw_{k}-\bw_{k+1}\|_G^2+2\langle\bw_{k+1}-\bw,F(\bw_{k+1})\rangle.
\end{align*}
\qed

\subsection*{7.3 Proof of Lemma 3}
To solve the case $\gamma=1$, we need further relax the inequality (15). Here, we focus on the term $\langle \br_{k+1}, B(\by_k-\by_{k+1})\rangle$.\\
Based on the optimality condition of the iteration of $\by$, we have following two inequalities:\\
\begin{equation*}
\left\{
\begin{aligned}
&\theta_2(\by)-\theta_2(\by_{k+1})+\langle \by-\by_{k+1}, T(\by_{k+1}-\by_k)-B^T\blambda_{k+1}+(1-\gamma)\beta B^T\br_{k+1}\rangle \geq 0 \\
&\theta_2(\by)-\theta_2(\by_{k})+\langle \by-\by_{k}, T(\by_{k}-\by_{k-1})-B^T\blambda_{k}+(1-\gamma)\beta B^T\br_{k}\rangle \geq 0
\end{aligned}
\right.
\ \ \forall \by \in \Y
\end{equation*}
Choose $\by$ to be $\by_k$ and $\by_{k+1}$ in two inequalities respectively\\
\begin{equation*}
\left\{
\begin{aligned}
&\theta_2(\by_k)-\theta_2(\by_{k+1})+\langle \by_k-\by_{k+1}, T(\by_{k+1}-\by_k)-B^T\blambda_{k+1}+(1-\gamma)\beta B^T\br_{k+1}\rangle \geq 0 \\
&\theta_2(\by_{k+1})-\theta_2(\by_{k})+\langle \by_{k+1}-\by_{k}, T(\by_{k}-\by_{k-1})-B^T\blambda_{k}+(1-\gamma)\beta B^T\br_{k}\rangle \geq 0
\end{aligned}
\right.
\end{equation*}
So, we have
\begin{align*}
&\langle B(\by_k-\by_{k+1}),-\blambda_{k+1}+(1-\gamma)\beta\br_{k+1}\rangle\geq \|\by_{k+1}-\by_k\|_T^2-(\theta_2(\by_k)-\theta_2(\by_{k+1}))\\
&\langle B(\by_k-\by_{k+1}),\blambda_k-(1-\gamma)\beta\br_{k}\rangle\geq-\langle\by_{k+1}-\by_k,T(\by_k-\by_{k-1})\rangle-(\theta_2(\by_{k+1})-\theta_2(\by_{k}))
\end{align*}
Combine these two inequalities together, we have
\begin{align*}
\langle B(\by_k-\by_{k+1}),\blambda_k-\blambda_{k+1}&+(1-\gamma)\beta\br_{k+1}\rangle-(1-\gamma)\beta\langle B(\by_k-\by_{k+1}),\br_k\rangle \\
&\geq \frac{1}{2}(\|\by_{k+1}-\by_k\|_T^2-\|\by_k-\by_{k-1}\|_T^2).
\end{align*}
Finally, using (6) we get
\begin{align*}
\langle \br_{k+1},B(\by_k-\by_{k+1})\rangle\geq&\frac{1-\gamma}{1+\alpha}\langle\br_k,B(\by_k-\by_{k+1})\rangle-\frac{\alpha}{1+\alpha}\|B(\by_k-\by_{k+1})\|^2\\
&+\frac{1}{2(1+\alpha)\beta}(\|\by_{k+1}-\by_k\|_T^2-\|\by_{k}-\by_{k-1})\|_T^2). \tag{16}
\end{align*}
So, we combine (15) and (16), then we have
\begin{align*}
&2P_1+(\|\bw_k-\bw\|_G^2+\frac{1-\alpha}{1+\alpha}\|\by_k-\by_{k-1}\|_T^2)-(\|\bw_{k+1}-\bw\|_G^2+\frac{1-\alpha}{1+\alpha}\|\by_{k+1}-\by_{k}\|_T^2)\\
&\geq \|\bu_k-\bu_{k+1}\|_P^2+(1-\alpha)\beta\|B(\by_k-\by_{k+1})\|^2-\frac{2(1-\alpha)\alpha}{1+\alpha}\beta\|B(\by_k-\by_{k+1})\|^2\\
&+(2-\alpha-\gamma)\beta\|\br_{k+1}\|^2+2(1-\gamma)\frac{1-\alpha}{1+\alpha}\beta\langle\br_k,B(\by_k-\by_{k+1})\rangle+2\langle\bw_{k+1}-\bw,F(\bw_{k+1})\rangle\\
&=\|\bu_k-\bu_{k+1}\|_P^2+\frac{(1-\alpha)^2}{1+\alpha}\beta\|B(\by_k-\by_{k+1})\|^2+(2-\alpha-\gamma)\beta\|\br_{k+1}\|^2\\
&+2(1-\gamma)\frac{1-\alpha}{1+\alpha}\beta\langle\br_k,B(\by_k-\by_{k+1})\rangle+2\langle\bw_{k+1}-\bw,F(\bw_{k+1})\rangle. \tag{17}
\end{align*}
Plug in $\gamma=1$ and define $c_2=\frac{1-\alpha}{1+\alpha}\in(0,1)$, we have
\begin{flalign*}
&2P_1+(\|\bw_k-\bw\|_G^2+c_2\|\by_k-\by_{k-1}\|_T^2)-(\|\bw_{k+1}-\bw\|_G^2+c_2\|\by_{k+1}-\by_{k}\|_T^2)\\
&\geq\|\bu_k-\bu_{k+1}\|_P^2+\frac{(1-\alpha)^2}{1+\alpha}\beta\|B(\by_k-\by_{k+1})\|^2+(1-\alpha)\beta\|\br_{k+1}\|^2+2\langle\bw_{k+1}-\bw,F(\bw_{k+1})\rangle\\
&=\|\bu_k-\bu_{k+1}\|_P^2+c_2((1-\alpha)\beta\|B(\by_k-\by_{k+1})\|^2+(1+\alpha)\beta\|\br_{k+1}\|^2)+2\langle\bw_{k+1}-\bw,F(\bw_{k+1})\rangle\\
&\stackrel{(14)}{=}\|\bu_k-\bu_{k+1}\|_P^2+c_2\|\bv_k-\bv_{k+1}\|_H^2+2\langle\bw_{k+1}-\bw,F(\bw_{k+1})\rangle\\
&\geq c_2\|\bw_k-\bw_{k+1}\|_G^2+2\langle\bw_{k+1}-\bw,F(\bw_{k+1})\rangle.
\end{flalign*}
\qed

\subsection*{7.4 Proof of Lemma 4}
In this lemma, we need further relax inequality (17). We use Cauchy-Schwarz inequality to deal with the term $\langle \br_k, B(\by_k-\by_{k+1})\rangle$. So, for any given $\delta>0$, we have
\begin{equation*}
-2\langle \br_k, B(\by_k-\by_{k+1})\rangle\geq-\delta\|\br_k\|^2-\frac{1}{\delta}\|B(\by_k-\by_{k+1})\|^2.
\end{equation*}
Plug in (17) then we have
\begin{flalign*}
2P_1&+(\|\bw_k-\bw\|_G^2+\frac{1-\alpha}{1+\alpha}\|\by_k-\by_{k-1}\|_T^2)-(\|\bw_{k+1}-\bw\|_G^2+\frac{1-\alpha}{1+\alpha}\|\by_{k+1}-\by_{k}\|_T^2)\\
&\geq \|\bu_k-\bu_{k+1}\|_P^2+(\frac{(1-\alpha)^2}{1+\alpha}-(\gamma-1)\frac{1-\alpha}{1+\alpha}\frac{1}{\delta})\beta\|B(\by_k-\by_{k+1})\|^2\\
&-\delta(\gamma-1)\frac{1-\alpha}{1+\alpha}\beta(\|\br_k\|^2-\|\br_{k+1}\|^2)+(2-\alpha-\gamma-\delta(\gamma-1)\frac{1-\alpha}{1+\alpha})\beta\|\br_{k+1}\|^2\\
&+2\langle\bw_{k+1}-\bw,F(\bw_{k+1})\rangle.
\end{flalign*}
Define $c_2=\frac{1-\alpha}{1+\alpha}$, $c_3=\delta(\gamma-1)\frac{1-\alpha}{1+\alpha}$ and $\tau=\frac{1-\alpha}{1+\alpha}\min\{1-\frac{\gamma-1}{1-\alpha}\frac{1}{\delta},\frac{\gamma-1}{\alpha+\gamma}(\frac{1+\alpha}{\gamma-1}-\frac{1+\alpha}{1-\alpha}-\delta)\}$, then transpose the corresponding terms and we will have
\begin{flalign*}
2P_1&+(\|\bw_k-\bw\|_G^2+c_2\|\by_k-\by_{k-1}\|_T^2+c_3\beta\|\br_k\|^2)\\
&-(\|\bw_{k+1}-\bw\|_G^2+c_2\|\by_{k+1}-\by_{k}\|_T^2+c_3\beta\|\br_{k+1}\|^2)\\
&\geq\|\bu_k-\bu_{k+1}\|_P^2+\frac{1-\alpha}{1+\alpha}(1-\frac{\gamma-1}{1-\alpha}\frac{1}{\delta})(1-\alpha)\beta\|B(\by_k-\by_{k+1})\|^2\\
&+(\gamma-1)\frac{1-\alpha}{1+\alpha}(\frac{1+\alpha}{\gamma-1}-\frac{1+\alpha}{1-\alpha}-\delta)\beta\|\br_{k+1}\|^2+2\langle\bw_{k+1}-\bw,F(\bw_{k+1})\rangle\\
&\geq\|\bu_k-\bu_{k+1}\|_P^2+\tau\|\bv_k-\bv_{k+1}\|_H^2+2\langle\bw_{k+1}-\bw,F(\bw_{k+1})\rangle\\
&\geq\tau\|\bw_k-\bw_{k+1}\|_G^2+2\langle\bw_{k+1}-\bw,F(\bw_{k+1})\rangle.
\end{flalign*}
Last, we need to verify our constants are reasonable and in (0,1).\\
First, if $\tau\in (0,1)$, we need following constraints
\begin{equation*}
\left\{
\begin{aligned}
&1-\frac{\gamma-1}{1-\alpha}\frac{1}{\delta} > 0\\
&\frac{1+\alpha}{\gamma-1}-\frac{1+\alpha}{1-\alpha}-\delta > 0
\end{aligned}
\right.
\ \ \ \Rightarrow \delta \in (\frac{\gamma-1}{1-\alpha},\frac{1+\alpha}{\gamma-1}-\frac{1+\alpha}{1-\alpha})
\end{equation*}
Also, it's easy to verify that if $\gamma>1$, then $\frac{\gamma-1}{1-\alpha}<\frac{1+\alpha}{\gamma-1}-\frac{1+\alpha}{1-\alpha}
\Longleftrightarrow \gamma\in (1,\frac{1-\alpha+\sqrt{(1+\alpha)^2+4(1-\alpha^2)}}{2})$. So this interval is reasonable under the condition of Lemma.\\
Second, we know $c_3=\delta(\gamma-1)\frac{1-\alpha}{1+\alpha}\in ((\gamma-1)\frac{1-\alpha}{1+\alpha}\frac{\gamma-1}{1-\alpha}, (\gamma-1)\frac{1-\alpha}{1+\alpha} (\frac{1+\alpha}{\gamma-1}-\frac{1+\alpha}{1-\alpha}))=(\frac{(\gamma-1)^2}{1+\alpha},\  2-\alpha-\gamma)\subset(0, 1)$ and $c_2=\frac{1-\alpha}{1+\alpha}\in(0, 1)$. So we have done the proof.\\
\qed

\subsection*{7.5 Proof of Theorem 1}
First, we combine Lemma 2, Lemma 3 and Lemma 4. We have if $\alpha\in [0, 1)$ and $\gamma \in (0,\frac{1-\alpha+\sqrt{(1+\alpha)^2+4(1-\alpha^2)}}{2})$, then there exist several constants $c_1, c_2, c_3\in (0,1)$ such that for any $\bw\in\Omega$, we have
\begin{align*}
2P_1&+(\|\bw_k-\bw\|_G^2+c_1\|\by_k-\by_{k-1}\|_T^2+c_2\beta\|\br_k\|^2)
-(\|\bw_{k+1}-\bw\|_G^2+c_1\|\by_{k+1}-\by_{k}\|_T^2+c_2\beta\|\br_{k+1}\|^2)\\
&\geq c_3\|\bw_k-\bw_{k+1}\|_G^2+2\langle\bw_{k+1}-\bw,F(\bw_{k+1})\rangle\geq2\langle\bw_{k+1}-\bw,F(\bw_{k+1})\rangle. \tag{18}
\end{align*}
Because of the monotonicity of $F(\bw)$, i.e.
\begin{equation*}
\langle \bw_{k+1}-\bw,F(\bw_{k+1})\rangle \geq \langle \bw_{k+1}-\bw,F(\bw)\rangle,
\end{equation*}
we have $\forall \bw\in\Omega$
\begin{align*}
\langle \bw_{k+1}-\bw,F(\bw)\rangle \leq P_1&+\frac{1}{2}(\|\bw_k-\bw\|_G^2+c_1\|\by_k-\by_{k-1}\|_T^2+c_2\beta\|\br_k\|^2)\\
&-\frac{1}{2}(\|\bw_{k+1}-\bw\|_G^2+c_1\|\by_{k+1}-\by_{k}\|_T^2+c_2\beta\|\br_{k+1}\|^2).
\end{align*}
Define \begin{align*}
f_k=\frac{1}{2}(\|\bw_k-\bw\|_G^2+c_1\|\by_k-\by_{k-1}\|_T^2+c_2\beta\|\br_k\|^2)
-\frac{1}{2}(\|\bw_{k+1}-\bw\|_G^2+c_1\|\by_{k+1}-\by_{k}\|_T^2+c_2\beta\|\br_{k+1}\|^2),
\end{align*}
we have
\begin{align*}
\langle \bw_{k+1}-\bw,F(\bw)\rangle \leq P_1+f_k. \tag{19}
\end{align*}
Based on inequality (19), we plug in $\bu=\bu^*$ and $\bw=\bw^*$ and have
\begin{align*}
\theta_1(\bx_k)+\theta_2(\by_{k+1})-\theta(\bu^*)+\langle\bw_{k+1}-\bw^*, F(\bw^*)\rangle
\leq \langle\bx-\bx_k,\delta_{k+1}\rangle+\frac{\eta_{k+1}}{2}\|\theta_1'(\bx_k,\xi_{k+1})\|^2+\frac{d_k}{2\eta_{k+1}}+f_k.
\end{align*}
We sum up the above inequality for all $k=1,2...t$, based on the definition of $\bar{\bu}_t$ and $\bar{\bw}_t$ and the convexity of $\theta$, we have 
\begin{align*}
&\theta(\bar{\bu}_t)-\theta(\bu^*)+\langle \bar{\bw}_t-\bw^*, F(\bw^*)\rangle\\
&\leq \frac{1}{t}\sum_{k=1}^{t}\langle\delta_{k+1}, \bx^*-\bx_k\rangle+\frac{1}{2t}\sum_{k=1}^{t}\eta_{k+1}\|\theta_1'(\bx_k,\xi_{k+1})\|^2+\frac{1}{2t}\sum_{k=1}^{t}\frac{d_k}{\eta_{k+1}}+\frac{1}{t}\sum_{k=1}^{t}f_k\\
&\leq \frac{1}{t}\sum_{k=1}^{t}\langle\delta_{k+1}, \bx^*-\bx_k\rangle+\frac{1}{2t}\sum_{k=1}^{t}\eta_{k+1}\|\theta_1'(\bx_k,\xi_{k+1})\|^2+\frac{\|\bx^*-\bx_1\|^2}{2t\eta_{t+1}}\\
&+\frac{1}{2t}(\|\bw_1-\bw^*\|_G^2+c_1\|\by_1-\by_0\|^2+c_2\beta\|\br_1\|^2). \tag{20}
\end{align*}
Take the expectation, under \textit{Assumption of (1) and (2)}, we get
\begin{equation*}
E[\theta(\bar{\bu}_t)-\theta(\bu^*)+\langle \bar{\bw}_t-\bw^*, F(\bw^*)\rangle]\leq \frac{N^2}{2t}\sum_{k=1}^{t}\eta_{k+1}+\frac{D_X^2}{2t\eta_{t+1}}+\frac{D}{2t}
\end{equation*}
where $D$ is a constant and $D_X$, $N$ as defined in notation and assumption (2).\\
So, if we set $\eta_k=Ck^{-p}$, then $\forall \epsilon>0$
\begin{align*}
\frac{1}{t}(\frac{N^2}{2}\sum_{k=1}^{t}\eta_{k+1}+\frac{D_X^2}{2\eta_{t+1}})&\leq\frac{D_X^2}{2C}\frac{(t+1)^p}{t}+\frac{N^2C}{2(1-p)}\frac{(t+1)^{1-p}}{t}\\
&\leq\frac{D_X^2}{2Ct^{1-p}}+\frac{N^2C}{2(1-p)t^p}+\epsilon\ \ \ \ (\text{if t large enough}). \tag{21}
\end{align*}
So, we have
\begin{align*}
E[\theta(\bar{\bu}_t)-\theta(\bu^*) + \langle\bar{\bw}_t-\bw^*, F(\bw^*)\rangle]
\leq \frac{D}{2t} + \frac{D_X^2}{2Ct^{1-p}} + \frac{N^2C}{2(1-p)t^{p}}+\epsilon.
\end{align*}
Moreover, for getting further result, we go back inequality (18). Note
\begin{align*}
\langle\bw_{k+1}-\bw,F(\bw_{k+1})\rangle\leq P_1+f_k.
\end{align*}
So, for any $\bw\in\Omega$ we have
\begin{align*}
\theta_1(\bx_k)+\theta_2(\by_{k+1})-\theta(\bu)+\langle\bw_{k+1}-\bw,F(\bw_{k+1})\rangle
\leq \langle\bx-\bx_k,\delta_{k+1}\rangle+\frac{\eta_{k+1}}{2}\|\theta_1'(\bx_k,\xi_{k+1})\|^2+\frac{d_k}{2\eta_{k+1}}+f_k.
\end{align*}
Then,
\begin{align*}
\theta(\bar{\bu}_t)-\theta(\bu)+\langle \bar{\bw}_t-\bw, F(\bar{\bw}_t)\rangle
\leq &\frac{1}{t}\sum_{k=1}^{t}\langle\delta_{k+1}, \bx-\bx_k\rangle+\frac{1}{2t}\sum_{k=1}^{t}\eta_{k+1}\|\theta_1'(\bx_k,\xi_{k+1})\|^2+\frac{\|\bx-\bx_1\|^2}{2t\eta_{t+1}}\\
&+\frac{1}{2t}(\|\bw_1-\bw\|_G^2+c_1\|\by_1-\by_0\|^2+c_2\beta\|\br_1\|^2).
\end{align*}
Note, this inequality hold for any $\bw\in\Omega$. So, $\forall \rho>0$, we set $\bw=(\bx^*, \by^*, \blambda)$ where $\blambda\in A=\{\blambda|\blambda\in\Lambda, \|\blambda\|_2\leq\rho\}$.\\
Also, we maximize left hand side and will have 
\begin{align*}
\max_{\blambda\in A}\{\theta(\bar{\bu}_t)-\theta(\bu)+\langle \bar{\bw}_t-\bw, F(\bar{\bw}_t)\rangle\}&=\max_{\blambda\in A}\{\theta(\bar{\bu}_t)-\theta(\bu^*)-\blambda^T(A\bar{\bx}_t+B\bar{\by}_t-\bb)\}\\
&=\theta(\bar{\bu}_t)-\theta(\bu^*)+\rho\|A\bar{\bx}_t+B\bar{\by}_t-\bb\|_2.
\end{align*}
So,
\begin{align*}
\theta(\bar{\bu}_t)-\theta(\bu^*)+\rho\|A\bar{\bx}_t+B\bar{\by}_t-\bb\|_2\leq &\frac{1}{t}\sum_{k=1}^{t}\langle\delta_{k+1}, \bx-\bx_k\rangle+\frac{1}{2t}\sum_{k=1}^{t}\eta_{k+1}\|\theta_1'(\bx_k,\xi_{k+1})\|^2+\frac{\|\bx-\bx_1\|^2}{2t\eta_{t+1}}\\
&+\frac{1}{2t}(\|\bw_1-\bw\|_G^2+c_1\|\by_1-\by_0\|^2+c_2\beta\|\br_1\|^2).
\end{align*}
Take expectation on both side, we have
\begin{align*}
E[\theta(\bar{\bu}_t)-\theta(\bu^*)+\rho\|A\bar{\bx}_t+B\bar{\by}_t-\bb\|_2]
\leq \frac{N^2}{2t}\sum_{k=1}^{t}\eta_{k+1}+\frac{D_X^2}{2t\eta_{t+1}}+\frac{D}{2t}.\tag{22}
\end{align*}
So, combine (21) with (22) and set $p=\frac{1}{2}$, we get final result
\begin{align*}
E[\theta(\bar{\bu}_t)-\theta(\bu^*)+\rho\|A\bar{\bx}_t+B\bar{\by}_t-\bb\|_2]=O(\frac{1}{\sqrt{t}}).
\end{align*}
\qed

\subsection*{7.6 Proof of Theorem 2}
Based on the definition of strong-convexity, we have
\begin{equation*}
\theta_1(\bx)-\theta_1(\bx_k)\geq\langle\theta_1'(\bx_k), \bx-\bx_k\rangle+\frac{\mu}{2}\|\bx-\bx_k\|^2.
\end{equation*}
As showed in (20), we have
\begin{align*}
\theta(\bar{\bu}_t)-\theta(\bu^*)+\langle \bar{\bw}_t-\bw^*, F(\bw^*)\rangle\leq& \frac{1}{t}\sum_{k=1}^{t}\langle\delta_{k+1}, \bx^*-\bx_k\rangle+\frac{1}{2t}\sum_{k=1}^{t}\eta_{k+1}\|\theta_1'(\bx_k,\xi_{k+1})\|^2\\
&+\frac{1}{t}\sum_{k=1}^{t}\left( (\frac{1}{2\eta_{k+1}}-\frac{\mu}{2})\|\bx_{k}-\bx^*\|^2-\frac{1}{2\eta_{k+1}}\|\bx_{k+1}-\bx^*\|^2\right)+\frac{D}{2t}.
\end{align*}
So, we take the expectation and set $\eta_k = \frac{1}{k\mu}$, then
\begin{flalign*}
&E[\theta(\bar{\bu}_t)-\theta(\bu^*)+\langle \bar{\bw}_t-\bw^*, F(\bw^*)\rangle]\\
&\leq \frac{D}{2t}+\frac{1}{t}E\bigg[\sum_{k=1}^{t}\left( (\frac{1}{2\eta_{k+1}}-\frac{\mu}{2})\|\bx_{k}-\bx^*\|^2-\frac{1}{2\eta_{k+1}}\|\bx_{k+1}-\bx^*\|^2\right)\bigg]\\
&+\frac{1}{2t}E[\sum_{k=1}^{t}\eta_{k+1}\|\theta_1'(\bx_k,\xi_{k+1})\|^2]\\
&\leq \frac{D}{2t}+\frac{N^2}{2t}\sum_{k=1}^{t}\frac{1}{\mu (k+1)}+\frac{1}{t}E\bigg[\sum_{k=1}^{t}\left(\frac{\mu k}{2}\|\bx_{k}-\bx^*\|^2-\frac{\mu(k+1)}{2}\|\bx_{k+1}-\bx^*\|^2\right)\bigg]\\
&\leq \frac{D+\mu D^2_X}{2t}+\frac{N^2\log(t+1)}{2\mu t}.
\end{flalign*}
So, $\forall\epsilon>0$, $\exists\  t$ large enough such that
\begin{flalign*}
E[\theta(\bar{\bu}_t)-\theta(\bu^*)+\langle \bar{\bw}_t-\bw^*, F(\bw^*)\rangle]
\leq \frac{D+\mu D^2_X}{2t}+\frac{N^2\log t}{2\mu t}+\epsilon.
\end{flalign*}
Moreover, we use similar way as in Theorem 1, then $\forall \rho>0$ and $\forall \bx\in\X$, we have
\begin{align*}
E[\theta(\bar{\bu}_t)-\theta(\bu^*)+\rho\|A\bar{\bx}_t+B\bar{\by}_t-\bb\|_2]
&\leq \frac{D}{2t}+\frac{1}{t}E\bigg[\sum_{k=1}^{t}\left( (\frac{1}{2\eta_{k+1}}-\frac{\mu}{2})\|\bx_{k}-\bx\|^2-\frac{1}{2\eta_{k+1}}\|\bx_{k+1}-\bx\|^2\right)\bigg]\\
&+\frac{1}{2t}E[\sum_{k=1}^{t}\eta_{k+1}\|\theta_1'(\bx_k,\xi_{k+1})\|^2]\\
&\stackrel{\eta_k = \frac{1}{k\mu}}{\leq} \frac{D+\mu D^2_X}{2t}+\frac{N^2\log(t+1)}{2\mu t}\\
&=O(\frac{\log t}{t}).
\end{align*}
\qed

\end{document}